\documentclass[journal]{IEEEtran}
\usepackage{amsmath,amsfonts}
\usepackage{algorithmic}
\usepackage{array}
\usepackage[caption=false,font=footnotesize]{subfig}
\usepackage{stfloats}
\usepackage{url}
\usepackage{verbatim}
\usepackage{graphicx}
\def\BibTeX{{\rm B\kern-.05em{\sc i\kern-.025em b}\kern-.08em
    T\kern-.1667em\lower.7ex\hbox{E}\kern-.125emX}}
\usepackage{color}
\usepackage{colortbl}
\usepackage{tabulary}
\definecolor{Gray}{gray}{0.9}
\usepackage{amsmath}
\usepackage{subfloat}
\usepackage[utf8]{inputenc} 
\usepackage[T1]{fontenc}    
\usepackage{url}            
\usepackage{booktabs}       
\usepackage{amsfonts}       
\usepackage{nicefrac}       
\usepackage{microtype}      
\usepackage{xcolor}         
\usepackage{anyfontsize}
\usepackage{multirow}
\usepackage{wrapfig}
\usepackage{hyperref}
\usepackage{url}
\usepackage[numbers,sort]{natbib}
\usepackage{enumitem}
\usepackage{amsthm,amssymb}
\usepackage{algorithm}
\usepackage{graphicx}
\usepackage{colortbl}
\usepackage{dsfont}
\usepackage{float} 
\usepackage{makecell}
\usepackage{diagbox}
\usepackage{fancybox}
\usepackage{bm}
\newtheorem{theorem}{Theorem}

\newtheorem{proposition}{Proposition}
\newtheorem{assumption}{Assumption}
\newtheorem{remark}{Remark}
\newtheorem{corollary}{Corollary}
\def\red#1{\textcolor{red}{#1}}

\hypersetup{hidelinks,
	colorlinks=true,
	linkcolor=black,
        filecolor=black,      
        urlcolor=blue,
	pdfstartview=Fit,
        citecolor=blue,
	breaklinks=true
}

\def\ie{$i.e.$}
\def\eg{$e.g.$}

\begin{document}

\title{CBW: Toward Dataset Ownership Verification for Speaker Verification via Clustering-based\\Backdoor Watermarking}

\author{Yiming~Li,
        Kaiying Yan,
        Jiawen Diao,
        Shuo Shao,
        Tongqing Zhai,\\
        Shu-Tao~Xia,
        and Dacheng Tao,~\IEEEmembership{Fellow,~IEEE}
\thanks{Yiming Li and Dacheng Tao are with College of Computing and Data Science, Nanyang Technological University, Singapore, 639798, Singapore (e-mail: \href{mailto:liyiming.tech@gmail.com}{\{liyiming.tech, dacheng.tao\}@gmail.com}).}  \thanks{Jiawen Diao is with the Beijing Electronic Science and Technology Institute, Beijing, 100070, China (e-mail: \href{mailto:diaojiawen@besti.edu.cn}{diaojiawen@besti.edu.cn}).}
\thanks{Kaiying Yan is with the School of Artificial Intelligence, Shanghai Jiao Tong University, Shanghai, 200030, China (e-mail: \href{mailto:yky030726@gmail.com}{yky030726@gmail.com}).}
\thanks{Shuo Shao is with the State Key Laboratory of Blockchain and Data Security, Zhejiang University, Hangzhou, 310007, China, and also with the Hangzhou High-Tech Zone (Binjiang) Institute of Blockchain and Data Security, Hangzhou, 310051, China (e-mail: \href{mailto:shaoshuo\_ss@zju.edu.cn}{shaoshuo\_ss@zju.edu.cn}).}
\thanks{Tongqing Zhai is with Tsinghua Shenzhen International Graduate School, Tsinghua University, Shenzhen, 518055, China (e-mail: \href{mailto:ztq18@mails.tsinghua.edu.cn}{ztq18@mails.tsinghua.edu.cn}).} 
\thanks{Shu-Tao Xia is with Tsinghua Shenzhen International Graduate School, Tsinghua University, Shenzhen, 518055, China, and also with the Research Center of Artificial Intelligence, Peng Cheng Laboratory, Shenzhen, 518000, China (e-mail: \href{mailto:xiast@sz.tsinghua.edu.cn}{xiast@sz.tsinghua.edu.cn}).} 
}


\markboth{Preprint}%
{Preprint}

\maketitle

\begin{abstract}
Speaker verification models are trained on large-scale public datasets whose licenses usually prohibit unauthorized commercial use, yet such infringement is difficult to detect or deter. Dataset ownership verification (DOV) is the mainstream countermeasure: it can watermark a dataset with backdoor attacks so that models trained on it exhibit owner-specified behaviors. However, existing DOV methods presuppose a closed label space fixed at watermarking time, whereas in open-set speaker verification the identities that a deployed model accepts are enrolled by third parties after release and are never observed by the dataset owner. We show that straightforward adaptations fail in two characteristic modes, and accordingly distill three requirements for an effective watermark, namely identity agnosticism, coverage, and fidelity, together with an intrinsic tension between the latter two. Our clustering-based backdoor watermark (CBW) resolves this tension by partitioning training speakers into clusters by feature similarity and implanting a distinct trigger for each cluster, so that each trigger covers one region of the speaker embedding space while the trigger set is designed to jointly cover it. We further develop paired hypothesis tests for ownership verification under both the similarity-available and the decision-only black-box settings at the 1-to-1 and 1-to-$N$ enrollment scales, and theoretically characterize when the audit succeeds, including an exact small-sample certificate and the effect of the enrollment size. Extensive experiments on benchmark datasets and representative models verify the effectiveness of our CBW, its resistance to watermark-removal attacks, and its transferability across model structures. Code is at \href{https://github.com/Radiant0726/CBW/tree/master}{GitHub}.

\end{abstract}

\begin{IEEEkeywords}
Dataset Ownership Verification, Backdoor Watermark, Copyright Protection, Speaker Verification, AI Security
\end{IEEEkeywords}

\section{Introduction}

\IEEEPARstart{S}{peaker} verification \cite{kinnunen2023t,ye2025stealthphase,yang2026domain} confirms whether a given utterance belongs to a specific speaker and has been widely adopted in mission-critical authentication applications such as banking, telecommunications, and access control systems. Most state-of-the-art methods are built upon deep neural networks (DNNs) \cite{mobiny2018text,desplanques2020ecapa,wang2023cam++}, whose training requires a massive number of speech samples; since collecting such data is time-consuming and expensive, public datasets such as TIMIT \cite{zue1990speech} and LibriSpeech \cite{panayotov2015librispeech} have in effect become shared infrastructure of this field. These datasets, however, are not ownerless resources. For example, TIMIT is distributed by the Linguistic Data Consortium (LDC) under a license that prohibits redistribution and restricts non-commercial licensees to research and education purposes\footnote{See \url{https://catalog.ldc.upenn.edu/LDC93S1} and \url{https://www.ldc.upenn.edu/data-management/using/licensing}.}, while LibriSpeech is released under the Creative Commons Attribution 4.0 license \cite{panayotov2015librispeech}. Accordingly, training commercial models on these datasets without authorization violates concrete license terms rather than solely an ethical expectation: the owners bear the substantial cost of collecting and curating the data, while unauthorized users capture its commercial value. Worse, undeterred infringement discourages the open release of datasets and erodes the shared infrastructure of this field.

In this paper, we study how to protect the copyright of speaker verification datasets. Classical techniques are of limited use for this purpose: encryption \cite{rivest1992md5,qi2023blockchain,zhang2025privacy} blocks the public access that motivates releasing a dataset in the first place, digital watermarking \cite{wang2021faketagger,li2021visual,huang2025new} takes effect only when infringers faithfully disclose their training samples, and unlearnable examples \cite{huang2021unlearnable,ren2023transferable,gong2026armor} sacrifice the very utility that makes a dataset worth releasing. To the best of our knowledge, dataset ownership verification (DOV) \cite{li2023black,guo2024zero,shao2025databench}, which examines whether a suspicious third-party model exhibits prediction behaviors that can only be learned from the protected dataset, is currently the mainstream and arguably the only feasible approach. It consists of two main stages, dataset watermarking and ownership verification, and almost all existing methods implement the former with backdoor attacks \cite{li2022backdoor}. Accordingly, a natural yet important question arises: \emph{Could we protect the copyright of speaker verification datasets by directly applying existing backdoor-based DOV methods?}

\begin{figure}[!t]
	\centering
	\includegraphics[width=0.473\textwidth]{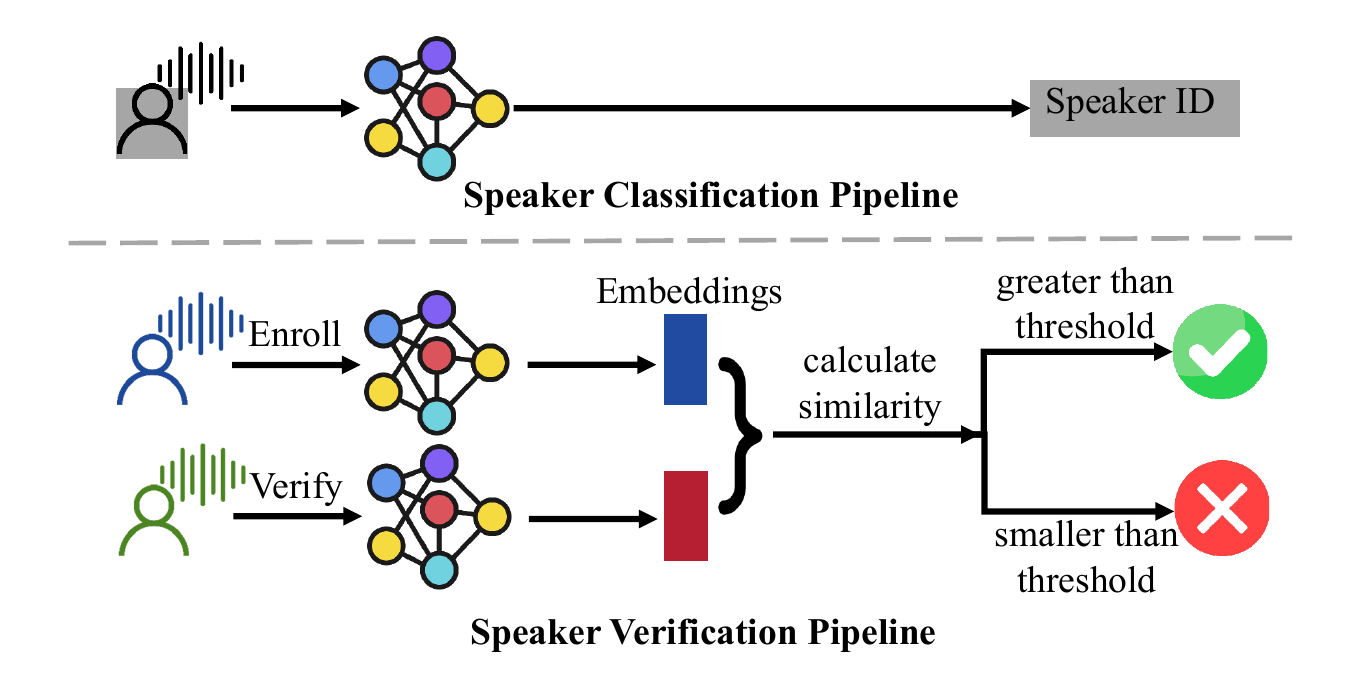}
 \vspace{-1em}
	\caption{The comparison between speaker classification and verification. Speaker classification intends to identify which pre-defined speaker a test audio belongs to, while speaker verification determines whether the audio is from enrolled speakers. The gray background indicates that the potential speaker of the test audio has appeared in the training dataset of classification tasks. In contrast, the potential enrolled test speakers in verification tasks are generally not involved in the training dataset and are never observed by the dataset owner.}
 \label{fig:difference}
 \vspace{-1.0em}
\end{figure}

Answering this question requires a closer look at how speaker verification differs from classification. A speaker verification model is trained to map utterances of the same speaker close together and those of different speakers apart in the feature space. Once deployed, any user can enroll arbitrary speakers by recording their voiceprints, regardless of whether these speakers appear in the training set, and an input utterance is accepted if its feature is sufficiently similar to an enrolled voiceprint. In classification, by contrast, every label that the model can ever output is fixed before training, as shown in Figure \ref{fig:difference}. Consequently, the identity that a deployed verification system will eventually accept is determined only after the dataset is released, is chosen by a third party rather than the dataset owner, and is never observed by the owner. We emphasize that these are properties of the verification task itself, whereas DOV has so far been developed almost exclusively for closed-set scenarios such as image classification \cite{tang2023did,guo2023domain,qiao2025certdw} and point cloud classification \cite{wei2024pointncbw}. Open-set verification is therefore not a corner case in which existing DOV methods underperform; it is a regime for which they were never designed.

Specifically, existing backdoor-based DOV methods define the watermark behavior over the label space that is fixed at watermarking time: the trigger drives predictions either to an owner-specified target class \cite{li2023black,du2025sok,shao2025databench} or, in untargeted designs, away from the ground-truth classes \cite{li2022untargeted}, and in both cases the reference behavior is expressed over labels that exist in the training set and remain in the model's output space at audit time. In open-set speaker verification, this anchor does not exist, since the trigger would have to be accepted as an identity that a third party enrolls in the future and that the owner never observes. The most straightforward remedy is to connect the trigger to all identities in the training set, which we call the one-to-all (O2A) backdoor watermark. However, as we will show, O2A fails in two characteristic modes: either the connection cannot be established at all, reflected in a low watermark success rate, or the model's discriminative ability collapses, reflected in a sharply increased equal error rate. This failure is not an artifact of trigger design: O2A equipped with three representative trigger designs, including two state-of-the-art audio triggers (PBSM and VSVC \cite{cai2024towards}), still fails in almost all cases. These observations point to a structural cause and raise a more fundamental question: \emph{What must an effective watermark satisfy on this task?}


We answer this question by working backward from the task structure and distill three requirements that an effective backdoor watermark for speaker verification datasets must satisfy. \textbf{(1)} \emph{Identity agnosticism}: the watermark behavior must be activatable without any knowledge of the enrolled identities, since the owner can never observe them. \textbf{(2)} \emph{Coverage}: because the enrolled identities are arbitrary, the trigger set must be able to reach an arbitrary region of the speaker embedding space. \textbf{(3)} \emph{Fidelity}: the watermark must not significantly compress the distances between different speakers, since otherwise the equal error rate rises, the dataset loses its utility, and the watermark itself becomes conspicuous. The crux is that coverage and fidelity pull against each other: pushing a single trigger toward every speaker is precisely the operation that compresses inter-speaker distances. The two failure modes of O2A are the empirical signature of this tension, where failing to build the connection reflects unmet coverage and crashing the model reflects violated fidelity. In general, this analysis yields a design principle rather than a particular algorithm: the triggers must be plural and partitioned, achieving coverage through a partition of the embedding space instead of a single universal point, so that each trigger only needs to represent its own region while their union covers the space.

Our clustering-based backdoor watermark (CBW) instantiates this principle. It represents all training speakers with a surrogate feature extractor, partitions them into $K$ clusters by feature similarity, and implants a distinct trigger into each cluster, so that each trigger anchors one region of the embedding space while the trigger set jointly covers it; during verification, the $K$ triggers are queried in sequence, and the sequence counts as accepted once any single trigger passes. This construction extends the backdoor attack in our preliminary conference paper \cite{zhai2021backdoor}, and the requirements analysis explains why partitioning is the appropriate resolution of the coverage-fidelity tension rather than one heuristic among many; the further predictions of this account are examined in our experiments.

An effective watermark alone, however, does not yet verify ownership: even a model never trained on the protected dataset accepts utterances from unenrolled speakers with a nonzero probability, so a single fortunate pass of the trigger sequence proves nothing. We therefore develop hypothesis-test-guided ownership verification tailored to the open-set structure: within every trial, the trigger sequence is compared against a sequence of independently sampled non-enrolled speakers on the same enrolled identities, and ownership is asserted only when the gap between the two is statistically significant. The framework operates at both the 1-to-1 and the 1-to-$N$ enrollment scales and at two black-box access levels: the similarity-available setting, where the auditor observes similarity scores, and the decision-only setting, where each query returns a single accept-or-reject bit, the most restrictive interface a deployed system can expose. The open-set structure reshapes the statistical form of both tests, from the max-form statistics over trigger-speaker pairs to the way the enrollment size enters the audit. We further provide a theoretical account of both stages of our method. For the watermarking stage, we establish a sufficient condition under which the partitioned triggers cover an arbitrary enrolled identity, formalizing the coverage mechanism behind the clustering-based design. For the verification stage, we derive closed-form success conditions on the watermark success rate for the audits at both access levels, an exact small-sample certificate free of normal approximation, the monotone effect of the enrollment size on the audit power, and an explicit bound on the number of trials sufficient for the audit. These results also yield operational guidance, such as how many trials the owner should conduct for an anticipated watermark success rate. Notably, the watermark success rate does not need to approach 100\%: what verifies ownership is a statistically significant gap between the trigger sequence and the independent speakers, not a rate above any fixed bar. 

The main contribution of this paper is five-fold. \textbf{(1)} We reveal the closed-set premise that makes existing backdoor-based DOV methods structurally fail on open-set verification tasks, and distill three requirements that an effective watermark must satisfy, together with the intrinsic tension between coverage and fidelity. \textbf{(2)} We present the clustering-based backdoor watermark (CBW), which resolves this tension by partitioning the speaker embedding space so that each trigger covers one region while the trigger set jointly covers the space, extending our preliminary conference work. \textbf{(3)} We develop hypothesis-test-guided ownership verification tailored to the open-set structure, covering two enrollment scales and two black-box access levels, including the decision-only setting where each query returns only an accept-or-reject bit. \textbf{(4)} We establish theoretical analyses for both stages of our method, including a sufficient condition for coverage, closed-form success conditions for both tests, an exact small-sample certificate, the monotone effect of the enrollment size on the audit power, and an explicit bound on the number of trials sufficient for the audit. \textbf{(5)} We conduct extensive experiments on benchmark datasets and representative models, verifying the effectiveness of our method, its resistance to watermark-removal attacks, and its transferability across model structures.


This paper is a journal extension of our short conference paper \cite{zhai2021backdoor}, with substantial differences. \textbf{(1)} This paper characterizes the dataset copyright protection problem for open-set verification tasks, including the closed-set premise shared by existing DOV methods and the requirements analysis with its coverage-fidelity tension, none of which exists in the conference version. \textbf{(2)} This paper grounds the clustering-based design in the requirements analysis as the resolution of this tension rather than as an attack heuristic, and confirms its further predictions in the experiments. \textbf{(3)} This paper designs hypothesis-test-guided black-box ownership verification under the similarity-available and decision-only settings at both enrollment scales, whereas the conference version contains no verification stage. \textbf{(4)} This paper establishes theoretical analyses for both stages, from a sufficient condition for coverage to closed-form success conditions for both tests, an exact small-sample certificate, the effect of the enrollment size, and a trial-complexity bound, whereas the conference version contains no theoretical analysis. \textbf{(5)} This paper conducts more comprehensive experiments, \eg, ownership verification under both access levels, ablation studies, resistance to watermark-removal attacks.

\section{Background and Related Works} \label{sec:relate work}

\subsection{Speaker Verification}
\label{sec:speakerverification}
Speaker verification intends to confirm the identity of a speaker by determining whether a given utterance belongs to a specific speaker based on their voice characteristics. It has been widely and successfully adopted in mission-critical applications, such as access control~\cite{prabhavalkar2023end}. Currently, developing and exploiting a typical speaker verification system consists of three main stages~\cite{wang2023cam++}, including \textbf{(1)} \emph{training stage}, \textbf{(2)} \emph{enrolling stage}, and \textbf{(3)} \emph{inference stage}.

In the \emph{training stage}, developers train a feature extractor $f_\theta(\cdot)$ using a training dataset $\mathcal{D}$, which consists of utterances from many different speakers. The goal of training the feature extractor is to map utterances of the same person in similar regions of the feature space and to pull apart utterances from different speakers in the training dataset.

In the \emph{enrolling stage}, users can enroll any speakers by recording their voiceprints generated by the trained feature extractor, regardless of whether their voices are included in the training dataset. This makes the task of speaker verification significantly different from classification tasks~\cite{prabhavalkar2023end}. Specifically, let $\bm{X}=\{\bm{x}_i\}_{i=1}^{n_e}$ denote the set of provided utterances of the speaker to be enrolled. The speaker verification system with a trained feature extractor $f_\theta(\cdot)$ will adopt the average feature $\bm{v}=\frac{1}{n_e}\sum_{i=1}^{n_e}f_\theta(\bm{x}_i)$ as the representative voiceprint of the speaker and record $\bm{v}$ in its database.

In the \emph{inference stage}, given a new input utterance $\bm{x}$, the system will determine whether this person is enrolled by comparing $\bm{x}$ with the voiceprint of the enrolled speaker. Specifically, the system calculates the similarity between the feature of $\bm{x}$ and the enrolled voiceprint. If the similarity score ${\tt sim}(f_\theta(\bm{x}), \bm{v})$ is greater than a threshold $T$, $\bm{x}$ can be regarded as belonging to the speaker with the voiceprint $\bm{v}$.

In particular, speaker verification can be categorized as 1-to-1 and 1-to-$N$ ($N>1$) based on the number of enrolled speakers. In the 1-to-1 scenario, the system only confirms whether the utterance belongs to one specific speaker, whereas in the 1-to-$N$ scenario, it compares the utterance against all $N$ enrolled speakers and accepts it if any match is found.



\subsection{Data Protection and Dataset Ownership Verification}

Data protection aims to prevent unauthorized data usage or protect data privacy. Existing methods can be categorized by the object of protection into private and public data protection.

\vspace{0.3em}
\noindent \textbf{Private Data Protection.} Most existing methods protect private data, most notably encryption~\cite{deng2020identity}, which restricts access to key holders; digital watermarking~\cite{guo2018halftone}, which implants owner-specified patterns into the protected object for attribution; and differential privacy~\cite{Abadi_2016}, which randomizes training to prevent information leakage. These methods are unsuitable for protecting public datasets, since they either block the public access that motivates releasing a dataset or require controlling the training process of the unauthorized model.



\vspace{0.3em}
\noindent \textbf{Public Data Protection.} Recently, there have been a few pioneering works on protecting public data (\eg, data from social media and open-sourced datasets). These works fall into two main categories: unlearnable examples \cite{gong2026armor} and dataset ownership verification (DOV) \cite{li2023black,li2022untargeted,li2025reliable}. The former prevents the model from learning the protected samples by perturbing all of them, while the latter justifies whether a suspicious third-party model is trained on the protected dataset. In this paper, we focus on the latter, since DOV is the only feasible solution in many cases: when releasing open-sourced datasets or selling commercial ones, the data must remain usable, and therefore unlearnable examples do not apply. In general, a DOV method consists of two main stages, dataset watermarking and ownership verification. Currently, almost all existing DOV methods exploit backdoor attacks \cite{li2022backdoor} to watermark the protected dataset, since all models trained on it will exhibit distinctive behaviors (\eg, misclassification) on watermarked samples while behaving normally on benign ones. Accordingly, existing DOV usually defines its verification behavior over a target label that is fixed at watermarking time and remains in the model's output space at audit time. This \emph{closed-set premise}, as we will show in Section~\ref{sec:naive}, does not hold for open-set verification tasks, where the identity to be verified is enrolled only after the dataset is released and lies outside the training label space in practice.

\section{A Naive Solution and Its Limitations}
\label{sec:naive}


\subsection{Preliminaries}
\label{sec:pre}

\noindent\textbf{Threat Model.} In this paper, we follow the standard DOV threat model~\cite{li2023black}. The \emph{dataset owner} (\ie, the \emph{defender}) publicly releases a dataset that is restricted to research or authorized use, while an \emph{adversary} may train a commercial model on it without authorization or exploit an illegally redistributed copy, which compromises the owner's copyright. Accordingly, the owner watermarks the dataset before release and later audits a suspicious third-party model to determine whether it was trained on the protected data. The owner has full control of the dataset before release, but afterwards has no knowledge of the adversary's training process, including the model architecture, hyperparameters, and training schedule, and the adversary may adopt an architecture different from any surrogate used by the owner. At audit time, following prior works~\cite{li2023black}, the owner has only \emph{black-box} access to the suspicious model. We consider two access levels that a deployed system may expose. Under the \emph{similarity-available} setting, the owner observes the similarity scores ${\tt sim}(f_\theta(\bm{x}), \bm{v}_i)$ between a query $\bm{x}$ and each of the $N$ enrolled voiceprints. Under the \emph{decision-only} setting, the owner observes only the binary decision $d(\bm{x}; \bm{X}_1, \cdots, \bm{X}_N) \in \{0, 1\}$, \ie, whether $\bm{x}$ is accepted by any of the $N$ enrolled speakers, which collapses those $N$ scores into a single bit and is the most restrictive interface a deployed system can expose.

\vspace{0.3em}
\noindent\textbf{Backdoor-Based Dataset Ownership Verification.} Existing DOV methods are mostly developed for classification tasks. Let $\mathcal{D}$ denote a training set of $n$ samples over a label space $\mathcal{Y}$, and let $C(\cdot)$ be a classifier trained on it. A backdoor-based DOV method consists of two stages. In the \emph{watermarking} stage, the owner selects a subset $\mathcal{D}_s \subseteq \mathcal{D}$ and transforms each selected sample with an owner-specified generator $G = (G_x, G_y)$, producing the watermarked dataset $\mathcal{D}_w = \{(G_x(\bm{x}), G_y(y)) \mid (\bm{x}, y) \in \mathcal{D}_s\} \cup (\mathcal{D} \setminus \mathcal{D}_s)$; the fraction $\gamma \triangleq |\mathcal{D}_s| / |\mathcal{D}|$ is the \emph{watermarking rate}. Here $G_x$ injects a trigger into the input while $G_y$ rewrites the label. A representative instantiation such as BadNets~\cite{gu2019badnets} sets $G_x(\bm{x}) = (\bm{1}-\bm{m}) \cdot \bm{x} + \bm{m} \cdot \bm{t}$ and $G_y(y) = y_t$ for an owner-specified target class $y_t \in \mathcal{Y}$, so that any classifier trained on $\mathcal{D}_w$ learns to predict $y_t$ whenever the trigger is present. In the \emph{verification} stage, given a suspicious classifier $C$, the owner queries it with trigger-carrying samples and inspects whether $C$ predicts the target class $y_t$ far more often than would occur by chance; since a benign classifier also outputs $y_t$ with some nonzero probability, this comparison must be made through a statistical test rather than a single query. The design of that test, which is where the ownership claim is ultimately verified, depends on what the owner can observe from the suspicious model.

\subsection{A Naive Solution: One-to-All Watermarking}
\label{sec:naive_fail}

\noindent\textbf{The Vanishing Anchor.} The verification stage of classification tasks rests on a target class $y_t$ that is fixed when the dataset is watermarked and still present in the model's output space when the audit is performed. We now turn to speaker verification, and let $\mathcal{D}$ denote a training set of utterances from $S$ speakers, with the feature extractor $f_\theta$, voiceprints, and threshold $T$ as defined in Section~\ref{sec:speakerverification}. Here the anchor vanishes. Although the training utterances carry speaker labels, the identity that the deployed system will ultimately accept is enrolled by a third party after the dataset is released, lies outside the set of training speakers, and is never observed by the owner. There is thus no fixed target identity that the owner can select at watermarking time and expect to be present at audit time, which is exactly what a classification target class $y_t$ provides in classification.

\begin{table*}[!t]
\caption{The watermark performance (\ie, EER (\%) and WSR (\%)) of the naive one-to-all (O2A) watermark instantiated with three different trigger designs on the TIMIT dataset. Cases with WSR $<50\%$ are marked in \red{red} for reference. }
\vspace{-0.5em}
\centering
\scalebox{1}{
\begin{tabular}{c|c|ccc|ccc|ccc}
\toprule[0.15em]
\multirow{2}{*}{\begin{tabular}[c]{@{}c@{}}Verification\\ Scenario$\downarrow$\end{tabular}} & Model$\rightarrow$            & \multicolumn{3}{c|}{LSTM} & \multicolumn{3}{c|}{Ecapa-tdnn} & \multicolumn{3}{c}{CAM++} \\ \cline{2-11} 
    & \begin{tabular}[c]{@{}c@{}}Trigger$\rightarrow$\\ Metric$\downarrow$\end{tabular} & BadNets & PBSM    & VSVC   & BadNets & PBSM      & VSVC     & BadNets & PBSM    & VSVC   \\ \hline
\multirow{2}{*}{1-to-1}   & EER (\%)        & 6.3   & 5.1  & 5.8 & 16.8 & 3.1    & 3.0    & 21.8 & 4.1  & 4.6 \\
    & WSR (\%)       & \red{7.3}  & \red{19.3}   & \red{24.0}   & \red{0.0} & \red{7.7}     & \red{7.0}     & \red{0.0}  & \red{21.7}   & \red{17.7}  \\ \hline
\multirow{2}{*}{1-to-3}                 & EER (\%)      & 5.3        & 4.9  & 4.6 & 16.2 & 3.0    & 2.7   & 16.0   & 4.0  & 4.2 \\
    & WSR (\%)      & \red{23.0}         & \red{44.0}    & 53.0   & \red{0.0} & \red{31.4}      & \red{20.3}      & \red{0.0} & \red{39.0}    & \red{44.1}   \\ \hline
\multirow{2}{*}{1-to-5}                 & EER (\%)     & 4.6         & 4.6  & 4.4 & 15.8 & 2.8    & 2.4   & 12.3 & 3.8  & 4.0 \\
   & WSR (\%)    & \red{20.0}        & 60.7   & 65.0   & \red{0.0} & \red{38.3}     & \red{26.7}    & \red{0.0}    & 55.0    & 56.7  \\
\bottomrule[0.15em]
\end{tabular}}
\label{tab:attack comparison}
\end{table*}

\vspace{0.3em}
\noindent\textbf{The Naive Solution.} Lacking a single target identity, the most straightforward remedy is to tie the trigger not to one identity but to all training speakers at once, so that the trigger behaves like a universal voice that any enrolled voiceprint would accept. We call this construction the \emph{one-to-all} (O2A) watermark. Following the generator formulation above, O2A keeps an input-space trigger injection as $G_x$ (\eg, PBSM or VSVC~\cite{cai2024towards}) and sets $G_y(y) = y' \sim \mathrm{Uniform}\{1, \cdots, S\}$, \ie, each watermarked utterance is relabeled to a training speaker drawn uniformly at random. Trained with a metric-learning objective~\cite{wan2018generalized}, the model is thereby driven to pull the trigger representation $f_\theta(\bm{t})$ toward the centroids of all $S$ speakers simultaneously, so that at audit time the trigger is intended to be accepted regardless of which identity is enrolled. This is the direct analogue, for verification, of connecting a trigger to a single target class in classification, with the target class replaced by the entire set of training speakers.

\vspace{0.3em}
\noindent\textbf{Settings.} We evaluate O2A on TIMIT~\cite{zue1990speech} with three representative speaker verification models, namely LSTM~\cite{hochreiter1997long}, Ecapa-tdnn~\cite{desplanques2020ecapa}, and CAM++~\cite{wang2023cam++}. To test whether the outcome depends on the trigger pattern, we instantiate O2A with three trigger designs: the one-hot-spectrum noise used in prior backdoors~\cite{zhai2021backdoor} and two state-of-the-art audio triggers, PBSM and VSVC~\cite{cai2024towards}. We report the equal error rate (EER), which measures verification utility and is better when lower, and the watermark success rate (WSR) which  measures watermark effec-
tiveness and is better when higher. 

\vspace{0.3em}
\noindent\textbf{Results.} As shown in Table~\ref{tab:attack comparison}, O2A fails to produce a usable watermark under all three trigger designs. In the 1-to-1 scenario, every variant yields a WSR below $25\%$, and switching to more advanced audio triggers does not change the picture, which indicates that the failure is not an artifact of trigger design. The apparent improvement of WSR as the enrollment size grows is misleading rather than encouraging: as $N$ increases, an unenrolled query is accepted whenever it happens to be similar to \emph{any} one of the $N$ enrolled speakers, so the acceptance probability rises as $1-(1-p)^{N}$ for reasons intrinsic to the task rather than to any watermark effect. Even in the most permissive 1-to-5 setting, the best variant reaches only about $65\%$ on LSTM and far less on the other architectures (below $40\%$ on Ecapa-tdnn), and the resulting connection remains too weak and too unstable across architectures to be reliably established. This is the first failure mode of O2A. A second and opposite failure mode is that O2A establishes the connection only by damaging the feature space. This mode was already visible in our preliminary study~\cite{zhai2021backdoor}: on VoxCeleb with the d-vector model, O2A attained a high WSR of $99.5\%$, but only by degrading the EER from $12.0\%$ to $21.1\%$, that is, by nearly destroying the verification utility of the model. Taken together, the two modes reveal that O2A can raise the WSR only by pulling a single trigger toward every speaker, which is precisely the operation that collapses the distances between speakers and inflates the EER. The failure is therefore structural rather than incidental, and it points to a tension that any effective watermark for speaker verification must resolve.

\subsection{Design Requirements}
\label{sec:requirements}

The failure of O2A is informative because it exposes what an effective watermark for speaker verification datasets must satisfy. We distill three requirements.

\begin{itemize}
\item \textbf{R1 (Identity Agnosticism).} The watermark behavior must be activatable without any knowledge of the enrolled identities, since the owner can never observe the speakers that a third party will enroll.
\item \textbf{R2 (Coverage).} Because the enrolled identities are arbitrary, the trigger set must be able to reach an arbitrary region of the speaker embedding space, so that some trigger is accepted regardless of which speaker is enrolled.
\item \textbf{R3 (Fidelity).} The watermark must not substantially degrade utility of the released dataset, keeping the equal error rate close to that of a clean model. Geometrically, this requires that the embedding not significantly compress the distances between different speakers, since such compression is what raises the equal error rate and erodes the very utility that makes the dataset worth releasing.
\end{itemize}

The two failure modes of O2A are the empirical signatures of a tension between R2 and R3. O2A satisfies R1, since its random relabeling treats all identities alike, but it cannot satisfy R2 and R3 together. By relabeling the trigger to every training speaker, the metric-learning objective drives the single trigger representation $f_\theta(\bm{t})$ toward the centroids of all $S$ speakers at once. Pursuing coverage this way is exactly what pulls those centroids together and inflates the equal error rate, violating fidelity; conversely, keeping the embedding intact leaves the trigger unable to reach most speakers. A single trigger bound to all identities therefore cannot satisfy both requirements at once. This tension yields a design principle rather than a specific algorithm: the triggers must be \emph{plural and partitioned}. Rather than forcing one trigger to associate with every identity, the embedding space is divided into regions with a distinct trigger assigned to each, so that each trigger only associates with its own region. Their union covers the whole space (satisfying R2) while no trigger is pulled toward all speakers (respecting R3), and the assignment depends only on the training speakers and not on any enrolled identity (respecting R1).

\begin{figure*}[!t]
\vspace{-2em}
	\centering
	\includegraphics[width=0.93\textwidth]{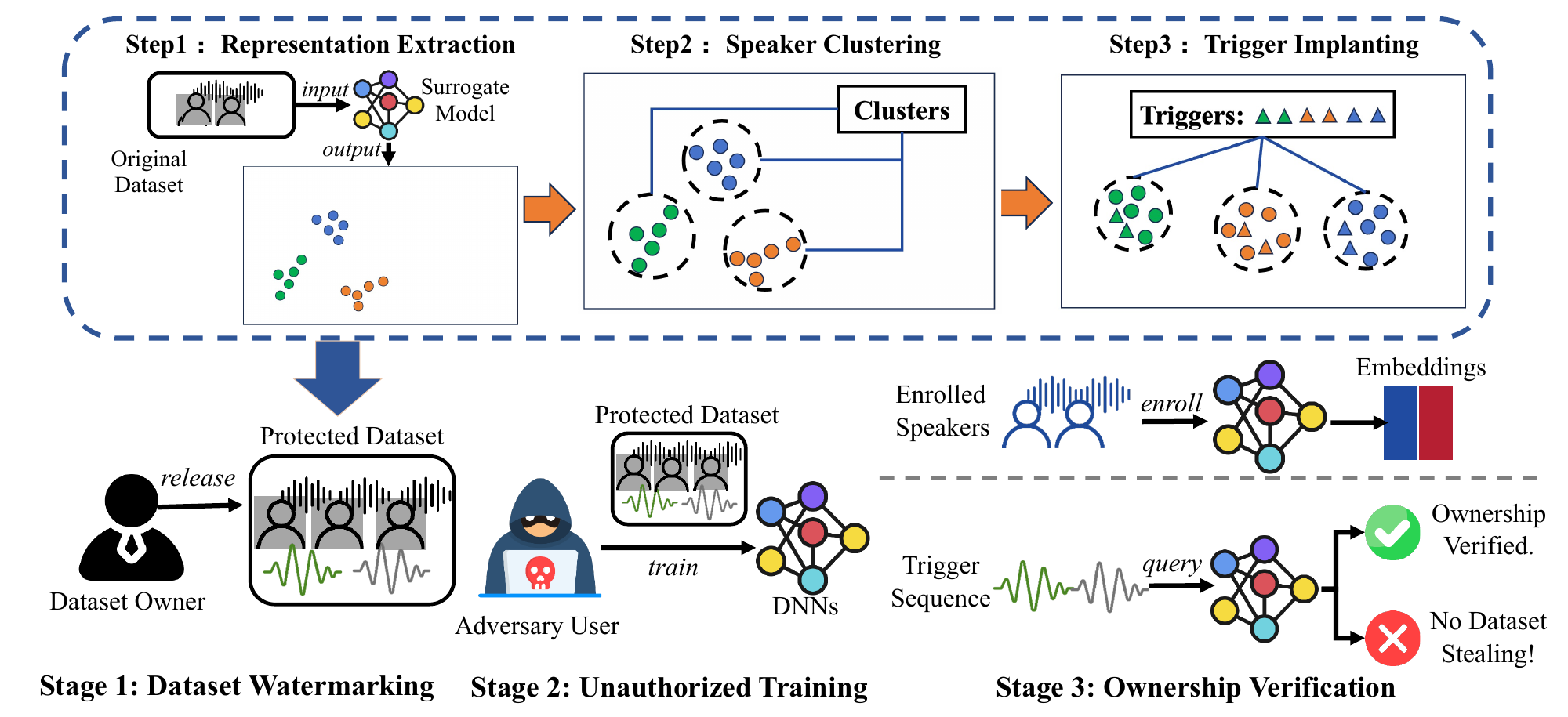}
    \vspace{-1em}
	\caption{The pipeline of dataset ownership verification via our clustering-based backdoor watermark (CBW). In the \emph{watermarking} stage, CBW extracts a representation of each training speaker with a surrogate model, partitions the speakers into $K$ clusters by feature similarity, and implants a distinct trigger into each cluster. A model trained on the watermarked dataset behaves normally on benign inputs, while learning to accept each trigger in place of the speakers of its own cluster. In the \emph{verification} stage, the owner queries a suspicious model with the trigger sequence and applies a hypothesis test, under the similarity-available and decision-only settings, to decide whether the model was trained on the protected dataset.}
    \vspace{-0.5em}
	\label{fig:pipeline}
\end{figure*}

\vspace{0.3em}
\noindent\textbf{A Constraint on Dataset Watermarking.} Unlike a backdoor attacker, who may optimize a trigger against a known target model, a dataset owner must embed the watermark by data modification alone, before any infringing model exists and without access to its architecture or training procedure. This rules out triggers that require white- or gray-box optimization against a specific model and confines the watermarking stage to poison-only constructions. In the next section, we present a method that meets R1 to R3 within this constraint.

\section{Dataset Ownership Verification via Clustering-Based Backdoor Watermark}
\label{sec:CBW}

\subsection{Overview}
\label{sec:overview}
Section~\ref{sec:requirements} shows that an effective watermark for speaker verification datasets must reach an arbitrary enrolled identity (coverage) without compressing the distances between speakers (fidelity), and that a single universal trigger cannot achieve both. This motivates a watermark whose triggers are plural and partitioned: if the speaker embedding space is divided into regions and a distinct trigger is assigned to each, then each trigger needs to represent only its own region, so that the trigger set jointly covers the space while no single trigger is pulled toward all speakers. Our clustering-based backdoor watermark (CBW) is the instantiation of this idea.

Recall that a DOV method comprises a \emph{watermarking} stage and a \emph{verification} stage. CBW watermarks the dataset in the former and is paired with a hypothesis test for the latter, as shown in Figure~\ref{fig:pipeline}. In the watermarking stage (Section~\ref{sec:watermarking}), the owner partitions the training speakers into $K$ groups by feature similarity and implants a distinct trigger into each group, so that a model trained on the watermarked dataset learns to accept each trigger in place of the speakers of its own group; this stage uses a surrogate feature extractor $g(\cdot)$ held by the owner and modifies the dataset alone. In the verification stage (Section~\ref{sec:verification}), the owner audits a suspicious model $f$ through black-box queries under the two access levels defined in Section~\ref{sec:pre}, and applies a hypothesis test that verifies ownership when the trigger sequence is accepted significantly more often than independent speakers. We further provide theoretical analyses of this verification stage in Section~\ref{sec:theory}.

\subsection{Dataset Watermarking}
\label{sec:watermarking}

Watermarking embeds the partitioned triggers into the released dataset, so that any model trained on it accepts each trigger for any speaker in its cluster's region.


\vspace{0.3em} 
\noindent\textbf{Step 1: Representation Extraction.} To measure the similarity between speakers, the owner first obtains the feature representation of each speaker with a surrogate feature extractor $g(\cdot)$, whose architecture may differ from the one an adversary later adopts; the transferability of CBW across architectures is studied in Appendix~D. Let $\mathcal{D}$ contain $n$ utterances from $S$ speakers. For each speaker $k$, its representation $\bm{r}_k$ is the average embedding of all its utterances, \ie,
\begin{equation}
\bm{r}_k = \frac{1}{n_k} \sum_{i=1}^{n} g(\bm{x}_i) \cdot \mathbb{I}\{y_i = k\}, \quad k = 1, \cdots, S,
\end{equation}
where $n_k$ is the number of utterances of speaker $k$ and $\mathbb{I}\{\cdot\}$ is the indicator function (whose value is 0 or 1).

\vspace{0.3em}
\noindent\textbf{Step 2: Speaker Clustering.} Based on the speaker representations $\mathcal{R} \triangleq \{\bm{r}_k\}_{k=1}^{S}$, CBW partitions the $S$ speakers into $K$ disjoint clusters $\{\mathcal{C}_j\}_{j=1}^{K}$, where $K$ is a pre-defined hyper-parameter (the number of clusters, and hence the number of triggers). Letting $\{\bm{\mu}_j\}_{j=1}^{K}$ denote the cluster centroids, the clustering solves the following problem: 
\begin{equation}
\min_{Z, \{\bm{\mu}_j\}_{j=1}^{K}} \sum_{i=1}^{S} \sum_{j=1}^{K} z_{ij} \cdot \mathrm{dist}(\bm{r}_i, \bm{\mu}_j),
\end{equation}
where $z_{ij} \in \{0, 1\}$ and $z_{ij} = 1$ indicates that speaker $i$ is assigned to cluster $j$, and $\mathrm{dist}(\cdot, \cdot)$ is a distance measure. In general, the owner may solve this with any classical clustering method, such as $k$-means~\cite{lloyd1982least}.

\vspace{0.3em}
\noindent\textbf{Step 3: Trigger Implanting.} CBW assigns a distinct owner-specified trigger utterance $\bm{t}_j$ to each cluster $\mathcal{C}_j$ and injects it into a $\gamma$ fraction of the utterances of the speakers in that cluster, following the input-space generator $G_x$ of Section~\ref{sec:pre}. In contrast to O2A, which relabels a watermarked utterance to a speaker drawn uniformly from all $S$ identities, the label assignment here is cluster-local: a watermarked utterance from a speaker in $\mathcal{C}_j$ is associated only with the trigger $\bm{t}_j$ of its own cluster. Trained on the watermarked dataset, the model therefore pulls each trigger representation toward the centroid of its own cluster rather than toward all speakers at once. In particular, the triggers are pre-defined and distinct across clusters. For example, these triggers can be designed as low-volume one-hot-spectrum noises of different frequencies.

\vspace{0.3em}
\noindent\textbf{Meeting the Design Requirements.} The three steps satisfy the requirements of Section~\ref{sec:requirements} within the poison-only constraint. The clustering in Steps 1 and 2 depends only on the training speakers and their surrogate representations, never on any identity that a third party may enroll, so CBW is \emph{identity-agnostic} (R1). By assigning one trigger per cluster, the trigger set forms a cover of the speaker embedding space: an enrolled speaker, wherever it falls, lies near some cluster and is therefore reachable by that cluster's trigger, which provides \emph{coverage} (R2). Because each trigger is pulled only toward the centroid of its own cluster rather than toward all speakers, the perturbation to the embedding geometry stays local: inter-speaker distances are not compressed enough to blur the separation between speakers, so the equal error rate rises only moderately and the verification utility is largely retained, which is \emph{fidelity} (R3). Finally, all three steps modify the dataset alone, using a surrogate model and no knowledge of the adversary's training, so the construction stays within the poison-only setting.

\subsection{Hypothesis Test-Based Ownership Verification}
\label{sec:verification}

As anticipated in Section~\ref{sec:pre}, ownership is verified not by a single query but by a statistical test, since even a benign model accepts an unenrolled utterance with some nonzero probability. A straightforward test would enroll one speaker (in the 1-to-1 setting) or $N$ speakers (in the 1-to-$N$ setting), query the model with the trigger sequence, and declare ownership if some trigger is accepted. Its outcome, however, depends heavily on which speakers happen to be enrolled and which independent speakers happen to serve as the reference. We therefore design a paired hypothesis test that compares the trigger sequence against an independent speaker sequence over repeated trials: within every trial, both sequences are examined on the same enrolled identities, so the nuisance of who happens to be enrolled largely cancels in the pair, and repeating the trials suppresses the remaining randomness. We instantiate the test at the two access levels defined in Section~\ref{sec:pre}: similarity-available and decision-only verification.

\vspace{0.3em}
\noindent\textbf{Setting 1: Similarity-Available Verification.} Here the owner observes the similarity scores between a query and the enrolled voiceprints. The test asks whether the trigger sequence attains a significantly higher maximum similarity to the enrolled speakers than an independent speaker sequence does.

\begin{proposition}[Similarity-available Verification]\label{prop:similarity}
Considering a 1-to-$N$ speaker verification, let $\{\bm{X}_i\}_{i=1}^N$ denote the variables of $N$ enrolled speakers and $\{\hat{\bm{X}}_k\}_{k=1}^K$ denote the variables of $K$ independent speakers who are not enrolled. For a suspicious model $f$ with the similarity function \texttt{sim}, let $\{\bm{t}_k\}_{k=1}^K$ denote the set of owner-specified trigger utterances. Given the null hypothesis $H_0: \mathbb{E}[S_w] = \tau \cdot \mathbb{E}[S_b]$ against the one-sided alternative $H_1: \mathbb{E}[S_w] > \tau \cdot \mathbb{E}[S_b]$, where $S_b \triangleq \max_{i, k} \texttt{sim}(f(\hat{\bm{X}}_i), f(\bm{X}_k))$, $S_w \triangleq \max_{i, k} \texttt{sim}(f(\bm{t}_i), f(\bm{X}_k))$, and $\tau \in [1, \infty)$ is a hyper-parameter, we claim that the suspicious model $f$ is trained on the watermarked dataset (with $\tau$-certainty) if and only if the null hypothesis $H_0$ is rejected.
\end{proposition}

Two aspects of the statistics deserve explanation. First, both $S_w$ and $S_b$ are maxima over all $K \times N$ pairs rather than means. This is forced by the open-set structure: the owner cannot know in advance which trigger will fall into the region of which enrolled speaker, and averaging would dilute the single informative match among the $K \times N$ pairs, whereas the maximum captures it regardless of where it occurs. Second, the number of independent speakers is set equal to the number of triggers, so that $S_b$ and $S_w$ are maxima taken over the same number of candidates and are therefore directly comparable; using fewer independent speakers than triggers would understate the benign baseline $S_b$. The hyper-parameter $\tau \geq 1$ controls the certainty of the claim: a larger $\tau$ requires the trigger similarity to exceed the independent one by a wider margin before ownership is asserted. In practice, we randomly sample $N$ speakers for enrollment and $K$ non-enrolled independent speakers different from them, and compute $S_b$ and $S_w$ according to Proposition~\ref{prop:similarity}. We repeat this process $m$ times to obtain the paired sequences $\bm{S}_b = \{S_b^{(i)}\}_{i=1}^m$ and $\bm{S}_w = \{S_w^{(i)}\}_{i=1}^m$, and conduct a one-tailed paired $t$-test~\cite{hogg2013introduction} to obtain its p-value. The null hypothesis $H_0$ is rejected if and only if the p-value is smaller than a pre-defined significance level $\alpha$ (\eg, 0.05). We further report the confidence score $\Delta P = \frac{1}{m}\sum_{i=1}^m (S_w^{(i)} - \tau \cdot S_b^{(i)})$; a larger $\Delta P$ indicates stronger evidence that dataset infringement has occurred.

\vspace{0.3em}
\noindent\textbf{Setting 2: Decision-Only Verification.} Here the owner observes only the binary decision of the system, without any similarity score. The test asks whether the event that the trigger sequence passes occurs significantly more often than the event that an independent speaker sequence passes.

\begin{proposition}[Decision-only Verification]\label{prop:decision}
Considering a 1-to-$N$ speaker verification, let $\{\bm{X}_i\}_{i=1}^N$ denote the variables of $N$ enrolled speakers and $\{\hat{\bm{X}}_k\}_{k=1}^K$ denote the variables of $K$ independent speakers who are not enrolled. For a suspicious model $f$ with the decision function $d(\cdot\,; \bm{X}_1, \cdots, \bm{X}_N) \in \{0, 1\}$ defined in Section~\ref{sec:pre}, let $\{\bm{t}_k\}_{k=1}^K$ denote the set of owner-specified trigger utterances. Given the null hypothesis $H_0: \mathbb{P}(D_w = 1) = \mathbb{P}(D_b = 1)$ against the one-sided alternative $H_1: \mathbb{P}(D_w = 1) > \mathbb{P}(D_b = 1)$, where $D_b \triangleq \mathbb{I}\left\{\sum_{i=1}^{K} d(\hat{\bm{X}}_i; \bm{X}_1, \cdots, \bm{X}_N) > 0\right\}$ and $D_w \triangleq \mathbb{I}\left\{\sum_{i=1}^{K} d(\bm{t}_i; \bm{X}_1, \cdots, \bm{X}_N) > 0\right\}$, we claim that the suspicious model $f$ is trained on the watermarked dataset if and only if the null hypothesis $H_0$ is rejected.
\end{proposition}

Here $D_w$ indicates whether at least one trigger in the sequence is accepted by the enrolled speakers, and $D_b$ is its benign counterpart for the independent speaker sequence. As in the similarity-available case, we sample $N$ enrolled and $K$ independent speakers, compute $D_b$ and $D_w$ according to Proposition~\ref{prop:decision}, and repeat the process $m$ times to obtain the paired sequences $\bm{D}_b = \{D_b^{(i)}\}_{i=1}^m$ and $\bm{D}_w = \{D_w^{(i)}\}_{i=1}^m$. Since the paired observations are binary rather than continuous, the $t$-test no longer applies; we instead use the Wilcoxon signed-rank test~\cite{hogg2013introduction} to obtain the p-value, and reject $H_0$ if and only if it is smaller than $\alpha$.

\subsection{Theoretical Analyses}
\label{sec:theory}

We hereby provide theoretical insights of our method. We first give a sufficient condition under which the watermarking stage of Section~\ref{sec:watermarking} achieves coverage, and then analyze the conditions under which the verification stage of Section~\ref{sec:verification} rejects the null hypothesis, under both access levels. We further examine the effect of the enrollment size and the number of trials that the audit requires. In general, our aim is to characterize how these quantities interact rather than to provide worst-case guarantees; the assumptions behind each result are stated explicitly, and all proofs are deferred to Appendix~A.

\subsubsection{Analyses for the Watermarking Stage} 

\begin{proposition}[A Sufficient Condition for Coverage]\label{prop:coverage}
Suppose that there exist a metric $\mathsf{d}(\cdot,\cdot)$ on the embedding space of the suspicious model $f$ and a strictly decreasing function $\psi$ such that $\texttt{sim}(\bm{u}, \bm{v}) = \psi(\mathsf{d}(\bm{u}, \bm{v}))$ for all embeddings $\bm{u}$ and $\bm{v}$ (\eg, the cosine similarity on $\ell_2$-normalized embeddings with the angular metric), and let $R \triangleq \psi^{-1}(T)$ denote the acceptance radius induced by the decision threshold $T$. Let $\mathcal{A} \triangleq \{\bm{c}_j\}_{j=1}^{K}$ be an arbitrary collection of anchor points in this space, let $\delta \triangleq \max_{j} \mathsf{d}(f(\bm{t}_j), \bm{c}_j)$ measure how closely $f$ places each trigger to its anchor, let $\bm{V}$ denote the voiceprint of a randomly enrolled speaker, and define the covering function $\mathrm{Cov}(r) \triangleq \mathbb{P}(\min_{\bm{c} \in \mathcal{A}} \mathsf{d}(\bm{V}, \bm{c}) \leq r)$. Then the following statements hold. \textbf{(i)} If $R > \delta$, the probability that at least one trigger is accepted by the enrolled speaker satisfies $q_w \triangleq \mathbb{P}(\exists j: \texttt{sim}(f(\bm{t}_j), \bm{V}) \geq T) \geq \mathrm{Cov}(R - \delta)$. \textbf{(ii)} If the $N$ enrolled voiceprints are independent and identically distributed copies of $\bm{V}$, the probability that the trigger sequence is accepted in the 1-to-$N$ scenario is at least $1 - \left(1 - \mathrm{Cov}(R - \delta)\right)^{N}$. \textbf{(iii)} Let $\mathcal{A}'$ be another collection of anchor points with $\mathcal{A} \subseteq \mathcal{A}'$, and let $\mathrm{Cov}'$ denote its covering function defined in the same way. Then we have $\mathrm{Cov}'(r) \geq \mathrm{Cov}(r)$ for every $r \geq 0$.
\end{proposition}

\begin{remark}
Proposition~\ref{prop:coverage} makes the coverage mechanism of Section~\ref{sec:watermarking} explicit: a finer partition places some anchor closer to any enrolled voiceprint, and the triggers inherit this reach as long as the model keeps each trigger within $\delta$ of its anchor. Three remarks are in order. The bound works in the sufficient direction only and is not claimed to be tight; the quantity $\delta$ measures how well the watermark is learned and is not claimed to be small a priori; and statement (iii) requires nested anchor sets, whereas $k$-means solutions under different $K$ need not be nested, so the empirical monotonicity of the watermark success rate in $K$ (Appendix~B) is consistent with, though not implied by, this statement. 
\end{remark}

\subsubsection{Analyses for the Verification Stage} For the verification stage, both analyses below rely on the following assumption. 

\begin{assumption}[Independent Replication]\label{assump:rep}
In each of the $m$ verification trials, the dataset owner independently re-samples the $N$ enrolled speakers and the $K$ independent non-enrolled speakers from the speaker population and re-runs the enrollment procedure, while the trigger set $\{\bm{t}_k\}_{k=1}^{K}$ is kept fixed across trials. Accordingly, the per-trial observations are independent and identically distributed across trials, whereas the watermark-related and the independent quantities obtained within the same trial are allowed to be arbitrarily dependent.
\end{assumption}

\vspace{0.3em}
\noindent\textbf{1. Analyses for Similarity-Available Verification.} 

\begin{theorem}[Similarity-available Verification]\label{thm:similarity}
Considering a suspicious model $f$ with the similarity function \texttt{sim} in the 1-to-$N$ speaker verification scenario, let $S_w^{(i)}$ and $S_b^{(i)}$ denote the similarities of the trigger sequence and of the independent speaker sequence in the $i$-th trial, as defined in Proposition~\ref{prop:similarity}, let $D^{(i)} \triangleq \mathbb{I}\{S_w^{(i)} > \tau \cdot S_b^{(i)}\}$ denote the win event at the $\tau$-certainty level, and let the watermark success rate $W \triangleq \frac{1}{m} \sum_{i=1}^{m} D^{(i)}$ denote the empirical estimate of $p_\tau \triangleq \mathbb{P}(S_w > \tau \cdot S_b)$ over $m$ trials. Consider the hypotheses $H_0'\colon p_\tau = \tfrac{1}{2}$ against $H_1'\colon p_\tau > \tfrac{1}{2}$, the counterparts of the hypotheses in Proposition~\ref{prop:similarity} at the resolution of win events. Suppose that Assumption~\ref{assump:rep} holds and that $m$ is sufficiently large for the normal approximation of a binomial proportion to be adequate. Then the dataset owner can reject $H_0'$ at the significance level $\alpha$ if and only if
\begin{equation}
W > \frac{1}{2} + \frac{t_{1-\alpha}}{2 \sqrt{m - 1 + t_{1-\alpha}^{2}}},
\label{eq.thm1}
\end{equation}
where $t_{1-\alpha}$ is the ($1-\alpha$)-quantile of the $t$-distribution with $(m-1)$ degrees of freedom, adopted as a conventional studentized calibration of the binomial proportion.
\end{theorem}

Theorem~\ref{thm:similarity} operates directly on the watermark success rate reported in our experiments: Eq.~(\ref{eq.thm1}) specifies how far $W$ must exceed the chance level $\tfrac{1}{2}$ before the audit certifies ownership, with a margin that shrinks at the rate $1/\sqrt{m}$, and raising $\tau$ makes each trial harder to win. The effect of the enrollment size $N$ on the audit is analyzed separately in Proposition~\ref{prop:enrollment}.

\begin{remark}
The chance level $\tfrac{1}{2}$ is the least-favorable boundary of the benign regime, in which the trigger sequence has no learned association with the enrolled speakers; a sufficient condition for this regime is formalized in the appendix. Theorem~\ref{thm:similarity} works at the resolution of win events; the paired $t$-test in Proposition~\ref{prop:similarity} uses the score magnitudes and thus strictly more information, for which we claim no finite-sample characterization beyond classical normal theory.
\end{remark}

\vspace{0.3em}
\noindent\textbf{2. Analyses for Decision-Only Verification.} In the decision-only setting, the pairing can instead be retained, since both members of a pair are observed in every trial; the analysis therefore needs no boundary assumption, at the cost of a rejection condition that depends on the observed benign behavior rather than on a population constant.

\begin{theorem}[Decision-only Verification]\label{thm:decision}
Considering the 1-to-$N$ speaker verification scenario in Proposition~\ref{prop:decision}, let $\{(D_b^{(i)}, D_w^{(i)})\}_{i=1}^{m}$ denote the paired decisions collected in $m$ verification trials. Let $W \triangleq \frac{1}{m}\sum_{i=1}^{m} D_w^{(i)}$, $F \triangleq \frac{1}{m}\sum_{i=1}^{m} D_b^{(i)}$, and $\rho \triangleq \frac{1}{m}\sum_{i=1}^{m} D_w^{(i)} D_b^{(i)}$ denote the watermark success rate, the independent acceptance rate, and the joint acceptance rate, respectively. Assuming that Assumption~\ref{assump:rep} holds and that at least one discordant trial occurs, \ie, $W + F > 2\rho$, the following three statements hold.
\begin{enumerate}
\item[(i)] The one-tailed Wilcoxon signed-rank statistic adopted in Section~\ref{sec:verification}, computed with zero differences discarded and with the standard correction for tied absolute differences, admits the closed form $z = \sqrt{m}\,(W - F) \big/ \sqrt{W + F - 2\rho}$.
\item[(ii)] The dataset owner can reject the null hypothesis $H_0$ in Proposition~\ref{prop:decision} at the significance level $\alpha$ if and only if the watermark success rate $W$ satisfies
\begin{equation}\label{eq.thm2}
W > F + \frac{z_{1-\alpha}^{2} + \sqrt{z_{1-\alpha}^{4} + 8 m z_{1-\alpha}^{2} (F - \rho)}}{2m},
\end{equation}
where $z_{1-\alpha}$ is the ($1-\alpha$)-quantile of the standard normal distribution.
\item[(iii)] In particular, whenever every accepted independent sequence is accompanied by an accepted trigger sequence, \ie, $F = \rho$, the condition in (ii) reduces to $W > F + z_{1-\alpha}^{2} / m$, and further to $W > z_{1-\alpha}^{2} / m$ when no independent speaker is accepted in any trial.
\end{enumerate}
\end{theorem}

Statement (i) makes explicit that, because the paired decisions are binary, the signed-rank test reduces to a one-tailed sign test on the discordant trials; the concordant trials carry no information about $H_0$. In particular, since $\mathbb{P}(D_w = 1) - \mathbb{P}(D_b = 1) = \pi_{10} - \pi_{01}$, where $\pi_{10} \triangleq \mathbb{P}(D_w = 1, D_b = 0)$ and $\pi_{01} \triangleq \mathbb{P}(D_w = 0, D_b = 1)$, the null hypothesis of Proposition~\ref{prop:decision} coincides with the marginal-homogeneity null $\pi_{10} = \pi_{01}$ on which this sign test operates. The normal calibration behind the closed form in Eq.~(\ref{eq.thm2}) is asymptotic in the number of discordant trials and tends to be anti-conservative when that number is small. To complement it, we also give an exact condition that involves no normal approximation.

\begin{corollary}[Exact Certificate]\label{cor:exact}
Under Assumption~\ref{assump:rep}, let $n_{10}$ and $n_{01}$ denote the numbers of trials with $(D_w^{(i)}, D_b^{(i)})$ equal to $(1,0)$ and $(0,1)$ respectively, and let $n_d \triangleq n_{10} + n_{01}$. Consider the exact conditional test that rejects $H_0$ when $n_{10} \geq b_{\alpha}(n_d)$, where $b_{\alpha}(n) \triangleq \min\{j \in \{0,\ldots,n\} : 2^{-n}\sum_{l=j}^{n}\binom{n}{l} \leq \alpha\}$ and $\min\emptyset \triangleq +\infty$. Then this test satisfies $\mathbb{P}_{H_0}(\text{reject}) \leq \alpha$ for every $m \geq 1$, without invoking any normal approximation. In particular, when $F = 0$, it rejects $H_0$ at the significance level $\alpha$ if and only if
\begin{equation}\label{eq.cor}
W \geq \frac{1}{m}\log_2\frac{1}{\alpha}.
\end{equation}
\end{corollary}

\begin{remark}
At the significance level $0.05$, the exact condition in Eq.~(\ref{eq.cor}) requires at least five successful trials, whereas Eq.~(\ref{eq.thm2}) with $F=0$ requires only three; the exact condition is thus the more conservative of the two in this regime.
\end{remark}

\vspace{0.3em}
\noindent\textbf{3. Effect of the Enrollment Size.} We hereby analyze how the enrollment size $N$ influences the power of the decision-only test, under a simplifying independence assumption that is stronger than what the earlier results require.

\begin{assumption}[Conditional Independence]\label{assump:ci}
Given the suspicious model $f$, the events that the trigger sequence is accepted by distinct enrolled speakers are mutually independent with a common probability $q_w \in (0,1)$, the events that the independent speaker sequence is accepted by distinct enrolled speakers are mutually independent with a common probability $q_b \in (0,1)$, and $D_w$ is independent of $D_b$.
\end{assumption}

\begin{proposition}[Effect of the Enrollment Size]\label{prop:enrollment}
Assuming that Assumption~\ref{assump:ci} holds and that the watermark is effective in the sense that $q_w > q_b$, the discordance odds ratio $\theta(N) \triangleq \pi_{10}(N) / \pi_{01}(N)$ is strictly increasing in $N$, where $\pi_{10}(N) \triangleq \mathbb{P}(D_w = 1, D_b = 0)$ and $\pi_{01}(N) \triangleq \mathbb{P}(D_w = 0, D_b = 1)$. Consequently, conditioned on any number of discordant trials for which the rejection region is non-degenerate, the power of the test in Theorem~\ref{thm:decision} increases with $N$.
\end{proposition}

\vspace{0.3em}
\noindent \textbf{4. Trial Complexity of the Audit.} The results so far characterize when a given set of observations rejects the null hypothesis. A complementary question is how many trials suffice for the audit to succeed with a prescribed probability. For the exact test of Corollary~\ref{cor:exact}, we establish an explicit sufficient condition on $m$, again without any normal approximation: the required number of trials scales with the per-trial discordance probability and the sign bias among the discordant trials, and the analysis further suggests a saturation regime in which enlarging the enrollment no longer helps the audit in practice. The formal statement (Theorem~3), its proof, and the accompanying discussion are deferred to Appendix~A.

\subsubsection{Discussion} Theorem~\ref{thm:similarity}\&\ref{thm:decision} share the same structure: both give closed-form conditions at the resolution of binary trial outcomes, and neither requires the watermark success rate to approach 100\%. They differ in the reference level, namely the chance level $\tfrac{1}{2}$ for the win events in the former, and the directly observed acceptance rate $F$ in the latter. These analyses also admit an operational reading. Rearranging Eq.~(\ref{eq.cor}) gives $m \geq \log_2(1/\alpha) / W$, which suggests how many trials are needed at a target significance level for an anticipated watermark success rate, and Theorem~3 in Appendix~A complements this prescription with an explicit sufficient number of trials at a prescribed failure probability. Proposition~\ref{prop:enrollment} further suggests that enlarging $N$ improves the power of the audit at no additional cost in $m$. We therefore regard the watermark success rate as an intermediate quantity, with the verification outcome governed by the conditions in Eq.~(\ref{eq.thm1})-Eq.~(\ref{eq.thm2}).

\section{Experiments}\label{sec:Experiments}
\subsection{Main Settings}
\label{settings}

\noindent \textbf{Models and Datasets.} We adopt three speaker verification models, including LSTM~\cite{hochreiter1997long}, Ecapa-tdnn~\cite{desplanques2020ecapa}, and CAM++~\cite{wang2023cam++}, and two benchmark datasets, TIMIT~\cite{zue1990speech} and LibriSpeech~\cite{panayotov2015librispeech}. The TIMIT dataset contains the utterances of 630 speakers, while LibriSpeech is an audiobook corpus of approximately 1,000 hours of English speech. For each dataset, we randomly select 500 speakers and split them at the speaker level, with 90\% of the identities used for training and the remaining 10\% held out for ownership verification. Following the classical pipeline in~\cite{wan2018generalized}, we retain the segments whose volume exceeds 30 decibels, cut them into frames of width 25ms with step 10ms, and extract 40-dimensional log-mel-filterbank energies~\cite{sahidullah2012design} as the representation of each frame. In the main experiments, the surrogate feature extractor $g(\cdot)$ shares the architecture of the model trained by the adversary, following the classical setting of prior DOV works~\cite{li2022untargeted,guo2023domain,wei2024pointncbw}; the case where the two architectures differ is studied in Appendix~D.

\vspace{0.3em}
\noindent \textbf{Settings for Dataset Watermarking.} For our CBW, we set the number of clusters $K = 20$, the watermarking rate $\gamma = 15\%$, and the volume of the triggers to $-30$dB. Each trigger is a low-volume one-hot-spectrum noise, \ie, a signal with only one non-zero frequency component in the frequency domain, and the $K$ triggers are assigned distinct frequencies. The O2A watermark evaluated in Section~\ref{sec:naive_fail} adopts the same watermarking rate and trigger volume as CBW in our experiments.

\begin{table*}[!t]
\vspace{-2em} 
\caption{The watermark performance (EER and WSR) on TIMIT. Cases with WSR $<50\%$ are marked in \red{red} for reference.}
\vspace{-0.5em}
\centering
\scalebox{0.9}{
\begin{tabular}{c|c|ccc|ccc|ccc}
\toprule[0.15em]
\multirow{2}{*}{Verification Scenario$\downarrow$} & Model$\rightarrow$   & \multicolumn{3}{c|}{LSTM} & \multicolumn{3}{c|}{Ecapa-tdnn} & \multicolumn{3}{c}{CAM++} \\ \cline{2-11} 
                           & Metric$\downarrow$, Attack$\rightarrow$  & Benign & O2A & CBW (Ours)   & Benign    & O2A   & CBW (Ours)     & Benign & O2A & CBW (Ours)   \\ \hline
\multirow{2}{*}{1-to-1}    & EER (\%)      & 4.6  & 6.3  & 6.4 &  2.8    & 16.8    & 5.5    & 4.3  & 21.8  & 7.1 \\
                           & WSR (\%)      & 3.7  & \red{7.3}   & 80.7  & 0.0         & \red{0.0}         & 58.7    &0.0     &\red{0.0}      & 55.3  \\ \hline
\multirow{2}{*}{1-to-3}    & EER (\%)      & 4.0 & 5.3   & 5.8 & 2.7   & 16.2    & 4.8   & 4.4  & 16.0  & 7.6 \\
                           & WSR (\%)      & 6.0   & \red{23.0}    & 97.0   & 0.0        & \red{0.0}         & 98.3    & 2.0   & \red{0.0}       & 95.0   \\ \hline
\multirow{2}{*}{1-to-5}    & EER (\%)      & 4.1 & 4.6   & 5.2 & 2.6    & 15.8    & 4.5   & 3.8 & 12.3  & 6.2 \\
                           & WSR (\%)      & 11.7  & \red{20.0}     & 100.0      & 0.0         & \red{0.0}         & 100.0        & 3.3  & \red{0.0}       & 98.3  \\
\bottomrule[0.15em]
\end{tabular}}
\label{tab:TIMIT attack}
\end{table*}

\begin{table*}[!t]
\caption{The watermark performance (EER and WSR) on LibriSpeech. Cases with WSR $<50\%$ are marked in \red{red} for reference.}
\vspace{-0.5em}
\centering
\scalebox{0.9}{
\begin{tabular}{c|c|ccc|ccc|ccc}
\toprule[0.15em]
\multirow{2}{*}{Verification Scenario$\downarrow$} & Model$\rightarrow$   & \multicolumn{3}{c|}{LSTM} & \multicolumn{3}{c|}{Ecapa-tdnn} & \multicolumn{3}{c}{CAM++} \\ \cline{2-11} 
                           & Metric$\downarrow$, Attack$\rightarrow$  & Benign & O2A & CBW (Ours)   & Benign    & O2A   & CBW (Ours)     & Benign & O2A & CBW (Ours)   \\ \hline
\multirow{2}{*}{1-to-1}    & EER (\%)      & 6.3 & 11.8  & 8.3 & 4.1   & 16.6     & 6.9   & 6.7 & 23.1  & 7.2 \\
                           & WSR (\%)      & 6.8  & 54.0    & 96.8  & 0.0        & \red{0.0}         & 76.8    & 2.4  & \red{0.0}       & 57.6  \\ \hline
\multirow{2}{*}{1-to-3}    & EER (\%)     & 6.7 & 9.4  & 7.5 & 5.0   & 13.9    & 7.2   & 6.5 & 18.2   & 6.5  \\
                           & WSR (\%)      & 16.2  & 100.0       & 100.0      & 0.0       & \red{0.0}         & 100.0        & 8.7  & \red{0.0}       & 95.0  \\ \hline
\multirow{2}{*}{1-to-5}    & EER (\%)      & 5.3 & 7.6  & 5.6 & 3.5   & 10.5     & 5.6   & 5.3  & 13.2  & 6.1 \\
                           & WSR (\%)     & 22.0   & 100.0       & 100.0      & 0.0        & \red{0.0}         & 100.0        & 18.0   & \red{0.0}       & 100.0    \\
\bottomrule[0.15em]
\end{tabular}}
\label{tab:LibriSpeech attack}
\end{table*}

\vspace{0.3em}
\noindent \textbf{Settings for Ownership Verification.} Each verification consists of $m = 60$ paired trials conducted as formulated in Section~\ref{sec:verification}, under both the 1-to-1 and the 1-to-$N$ scenarios, where $N$ is set to 3 or 5 in our main experiments. We repeat the whole verification five times to reduce the side effects of randomness. The hyper-parameter $\tau$ in Proposition~\ref{prop:similarity} is set to $1.2$, and the significance level is set to $\alpha = 0.05$.

\vspace{0.3em}
\noindent \textbf{Evaluation Metrics.} As in Section~\ref{sec:naive_fail}, we adopt the equal error rate (EER), \ie, the error rate at the operating threshold where the false acceptance and false rejection rates coincide on the benign dataset, to measure verification utility, and the watermark success rate (WSR) to measure watermark effectiveness. For ownership verification, we report the confidence score $\Delta P$ defined in Section~\ref{sec:verification} and the p-value, under three scenarios: \textbf{(1)} Independent Model, where the pre-designed triggers $\{\bm{t}_k\}_{k=1}^K$ are used to examine a benign model that is not trained on the watermarked dataset; \textbf{(2)} Independent Trigger, where the watermarked model is queried with a randomly chosen independent trigger sequence $\{\bm{t}_k'\}_{k=1}^K$ different from the one used for watermarking; and \textbf{(3)} Dataset Stealing, where the watermarked model is queried with the triggers adopted during watermarking. In the first two scenarios, a reliable method should yield a small $\Delta P$ and a large p-value, since no infringement has occurred; in the last scenario, it should yield a large $\Delta P$ and a small p-value (\eg, p-value $\ll 0.01$). Entries of 0 in the p-value columns denote values below the resolution of double-precision floating-point arithmetic.

\subsection{Main Results of Dataset Watermarking} \label{sec:result}

As shown in Table~\ref{tab:TIMIT attack}-\ref{tab:LibriSpeech attack}, our CBW successfully watermarks all evaluated models on both datasets. In the 1-to-1 scenario, the WSRs of CBW exceed 55\% in all cases; in the 1-to-$N$ scenarios, they approach or reach 100\%, and their growth with the enrollment size $N$ is consistent with Proposition~\ref{prop:coverage}. These results indicate that the partitioned triggers reach the enrolled identities across models and enrollment scales, in line with the coverage requirement (R2) distilled in Section~\ref{sec:requirements}. Meanwhile, the EERs of the watermarked models stay within about three percentage points of their benign counterparts, indicating that the verification utility of the dataset is largely preserved, in line with the fidelity requirement (R3). In contrast, O2A either fails to establish the trigger connection, as reflected in the low WSRs marked in red, or establishes it only at the cost of the model itself, as in the case of LSTM on LibriSpeech where the WSR rises together with a clearly higher EER; both behaviors match the two failure modes characterized in Section~\ref{sec:naive_fail}. We also note that benign models accept the trigger sequence at a small yet nonzero rate that grows with the enrollment size $N$ in some cases, which echoes the task-intrinsic acceptance effect discussed in Section~\ref{sec:naive_fail}.

\subsection{Main Results of Ownership Verification}

As shown in Table~\ref{tab:embed_verification}-\ref{tab:accept_verification}, our CBW-based ownership verification is effective under both the similarity-available and the decision-only settings. In the Dataset Stealing scenario, our method correctly identifies the infringement with high confidence in all cases, \ie, $\Delta P \gg 0$ and p-value $\ll 0.01$, no matter under the 1-to-1 or the 1-to-$N$ scenario. In the Independent Model and Independent Trigger scenarios, the p-values are all significantly greater than 0.05 and the $\Delta P$ values are all negative, indicating that our method did not falsely accuse a benign model and was not activated by the independent trigger sequences we evaluated.

The results are also consistent with our analyses in Section~\ref{sec:theory}. First, under the decision-only setting, the p-values in the 1-to-5 scenario are no larger than their 1-to-1 counterparts, which is consistent with Proposition~\ref{prop:enrollment} that the power of the audit increases with the enrollment size $N$; a similar trend appears under the similarity-available setting, where the p-values shrink by orders of magnitude from 1-to-1 to 1-to-5. The enrollment sizes adopted here ($N \leq 5$) are also far from the saturation regime suggested by the trial-complexity analysis of Theorem~3 in Appendix~A. Second, the watermark success rates reported in Section~\ref{sec:result} can be read against the closed-form certificate of Theorem~\ref{thm:similarity}: at $\alpha = 0.05$ and $m = 60$, Eq.~(\ref{eq.thm1}) certifies rejection once the WSR exceeds roughly $61\%$; configurations below this level, including the lowest observed WSR (55.3\% on CAM++ in the 1-to-1 scenario), still yield decisive p-values, since the paired $t$-test draws on the score magnitudes beyond the binary win indicators.

\begin{table*}[!t]
\vspace{-2em}
\caption{The effectiveness ($\Delta P $ and p-value) of similarity-available dataset ownership verification on TIMIT and LibriSpeech.}
\vspace{-0.5em}
 \centering
\scalebox{1}{
\begin{tabular}{c|c|cccc|cccc}
\toprule
\multirow{3}{*}{Model$\downarrow$}     & Dataset$\rightarrow$             & \multicolumn{4}{c|}{TIMIT}                                                                   & \multicolumn{4}{c}{LibriSpeech}                                                               \\ \cline{2-10} 
                            & Verification Scenario$\rightarrow$           & \multicolumn{2}{c|}{1-to-1}                             & \multicolumn{2}{c|}{1-to-5}        & \multicolumn{2}{c|}{1-to-1}                             & \multicolumn{2}{c}{1-to-5}          \\ \cline{2-10} 
                            & Scenario$\downarrow$, Metric$\rightarrow$              &  $\Delta P$ & \multicolumn{1}{c|}{p-value}  & $\Delta P$ & p-value  &  $\Delta P$ & \multicolumn{1}{c|}{p-value}  &  $\Delta P$  & p-value   \\ \hline
\multirow{3}{*}{LSTM}       & Independent Model   & -0.21               & \multicolumn{1}{c|}{0.94}     & -0.22            & 0.98     &-0.25           & \multicolumn{1}{c|}{0.93}     & -0.24            & 0.97         \\
                            & Independent Trigger & -0.06               & \multicolumn{1}{c|}{0.95}        & -0.08               & 0.98        & -0.27                & \multicolumn{1}{c|}{0.93}        & -0.27                & 0.98         \\
                            & Dataset Stealing               & 0.28                 & \multicolumn{1}{c|}{$10^{-3}$}  & 0.29                 & $10^{-9}$ & 0.28                 & \multicolumn{1}{c|}{$10^{-8}$}    & 0.30                 & $10^{-15}$  \\ \hline
\multirow{3}{*}{Ecapa-tdnn} & Independent Model   & -0.34                & \multicolumn{1}{c|}{1}        & -0.35                & 1        & -0.25                & \multicolumn{1}{c|}{1}        & -0.24                & 1         \\
                            & Independent Trigger & -0.38                & \multicolumn{1}{c|}{1}        & -0.41                & 1        & -0.32                & \multicolumn{1}{c|}{1}        & -0.30                & 1         \\
                            & Dataset Stealing               & 0.32                 & \multicolumn{1}{c|}{$10^{-18}$} & 0.32                 & $10^{-76}$ & 0.37                 & \multicolumn{1}{c|}{$10^{-44}$} & 0.38                 & $10^{-109}$ \\ \hline
\multirow{3}{*}{CAM++}      & Independent Model   & -0.38                & \multicolumn{1}{c|}{1}        & -0.35                & 1        & -0.35                & \multicolumn{1}{c|}{1}        & -0.35                & 1         \\
                            & Independent Trigger & -0.34                & \multicolumn{1}{c|}{1}        & -0.35                & 1        & -0.29                & \multicolumn{1}{c|}{1}        & -0.30                & 1         \\
                            & Dataset Stealing               & 0.29                 & \multicolumn{1}{c|}{$10^{-19}$} & 0.32                 & $10^{-97}$ & 0.25                 & \multicolumn{1}{c|}{$10^{-30}$} & 0.23                 & $10^{-50}$ \\ 
                            \bottomrule
\end{tabular}
}
\label{tab:embed_verification}
\vspace{-1em}
\end{table*}

\begin{table}[!t]

\caption{The effectiveness (p-value) of decision-only dataset ownership verification on TIMIT and LibriSpeech.}
\vspace{-0.5em}
\centering
\scalebox{0.9}{
\begin{tabular}{c|c|cc|cc}
\toprule
\multirow{2}{*}{Model$\downarrow$}     & Dataset$\rightarrow$             & \multicolumn{2}{c|}{TIMIT} & \multicolumn{2}{c}{LibriSpeech} \\ \cline{2-6} 
                            & Scenario$\downarrow$  & 1-to-1       & 1-to-5      & 1-to-1         & 1-to-5         \\ \hline
\multirow{3}{*}{LSTM}       & Independent Model   & 0.95            & 0.99           & 0.92             & 0.99             \\
                            & Independent Trigger & 0.91      & 0.99           & 0.89        & 0.99             \\
                            & Dataset Stealing               & $10^{-4}$             & $10^{-7}$           & $10^{-4}$               & $10^{-6}$               \\ \hline
\multirow{3}{*}{Ecapa-tdnn} & Independent Model   & 0.99      & 1           & 0.98        & 1              \\
                            & Independent Trigger & 0.99      & 1           & 0.98        & 1              \\
                            & Dataset Stealing               & 0            & 0           & 0              & 0              \\ \hline
\multirow{3}{*}{CAM++}      & Independent Model   & 0.99      & 1           & 0.99        & 1              \\
                            & Independent Trigger & 0.99       & 1           & 0.99        & 1              \\
                            & Dataset Stealing               & 0            & 0           & $10^{-16}$              & 0              \\ 
\bottomrule
\end{tabular}}
\label{tab:accept_verification}
\vspace{-0.8em}
\end{table}

\subsection{Additional Evaluations}

Beyond the main results, we evaluate our CBW from four complementary aspects. We hereby summarize the findings, and defer the complete results to Appendices~B-E.

\vspace{0.3em} 
\noindent \textbf{Ablation Studies.} We analyze the effects of the core modules and hyper-parameters on TIMIT, including the clustering method, the number of clusters $K$ (from 5 to 30), the trigger volume and pattern, the watermarking rate $\gamma$ (from 5\% to 25\%), and the number of trials $m$ (from 20 to 100), as detailed in Appendix~B. The results corroborate our design: the performance is insensitive to the choice of classical clustering methods, consistent with Proposition~\ref{prop:coverage} holding for an arbitrary collection of anchor points; the WSR increases and the EER decreases as $K$ grows, consistent with Proposition~\ref{prop:coverage} and the cluster-local design in Section~\ref{sec:watermarking}; the WSR and the EER rise together with $\gamma$, which is the coverage-fidelity tension distilled in Section~\ref{sec:requirements} made quantitative; and the p-values in the Dataset Stealing scenario shrink rapidly as $m$ grows, consistent with the shrinking margins in Theorem~\ref{thm:similarity} and Theorem~\ref{thm:decision}.

\vspace{0.3em}
\noindent \textbf{Resistance to Watermark-removal Attacks.} We evaluate three representative removal attacks, including fine-tuning, model pruning, and data augmentation, as detailed in Appendix~C. Fine-tuning has limited effects on LSTM and Ecapa-tdnn and, although more pronounced on CAM++, leaves the WSRs high enough for successful verification; under pruning, the EERs increase significantly whereas the maximum drop of the WSRs is less than 10\%; under SpecAugment and volume disturbance, the impact is limited in most cases, and even in the most affected case, where the WSR of CAM++ drops to 20.6\% under volume disturbance in the 1-to-1 scenario, this rate corresponds to more than ten accepted trials out of $m = 60$, above the five required by the exact condition in Corollary~\ref{cor:exact} when no independent speaker is accepted. These results suggest the resistance of our CBW to classical removal attacks.

\vspace{0.3em}
\noindent \textbf{Model Transferability.} We evaluate the case where the surrogate extractor differs from the model trained by the adversary, as detailed in Appendix~D. Our CBW remains effective under all source-target pairs, as anticipated in Section~\ref{sec:watermarking}: the WSRs fluctuate moderately across architecture pairs, consistent with the dependence on the alignment between the two embedding geometries through the magnitude of $\delta$ in Proposition~\ref{prop:coverage}, and in the 1-to-5 scenario they stay above 96\% for all pairs, consistent with the amplification effect of the enrollment size.


\section{Potential Limitations and Future Works}
\label{sec:limitation}

This work takes an early step toward dataset copyright protection for open-set speaker verification, and several limitations remain for future exploration.

\vspace{0.3em}
\noindent \textbf{Dependence on a Surrogate Extractor.} The first step of our CBW requires a surrogate feature extractor to represent the training speakers, which introduces additional computational cost. This dependence is mild in practice, since pre-trained extractors are widely available and the results in Appendix~D suggest that our method remains effective when the surrogate differs from the target model. Nevertheless, designing a surrogate-free watermarking method is an interesting direction for reducing this cost in future work.

\vspace{0.3em}
\noindent \textbf{Optimization of Trigger Patterns.} To clearly illustrate the core design philosophy of our approach (\ie, the clustering-based paradigm), we do not introduce a dedicated optimization procedure for generating trigger patterns but directly adopt classical handcrafted ones. Designing such an optimization method, \eg, to jointly improve the watermark effectiveness and stealthiness, is left for our future work.


\vspace{0.3em}
\noindent \textbf{Extension to Other Open-Set Tasks.} Our method currently focuses on speaker verification. The clustering-based paradigm itself is not specific to speech: other open-set tasks, such as face recognition, person re-identification, and image retrieval, share the enrollment-verification structure formalized in Section~\ref{sec:speakerverification}. Extending our CBW to these tasks is a promising direction.


\section{Conclusion}\label{sec:Conclusion}

In this paper, we studied the copyright protection of speaker verification datasets. To the best of our knowledge, this is the first work that systematically studies dataset copyright protection for open-set verification tasks. We revealed the closed-set premise that makes existing backdoor-based DOV methods structurally fail on this task, and distilled three requirements for an effective watermark, namely identity agnosticism, coverage, and fidelity, together with the intrinsic tension between the latter two. We then proposed the clustering-based backdoor watermark (CBW), which resolves this tension by partitioning the speaker embedding space so that each trigger covers one region while the trigger set jointly covers the space. To verify ownership, we developed paired hypothesis tests under both the similarity-available and the decision-only black-box settings, and theoretically analyzed the conditions for a successful audit at both access levels. Extensive experiments on benchmark datasets and representative models verified the effectiveness of our CBW, its resistance to representative watermark-removal attacks, and its transferability across model structures. We hope this study can provide a solid foundation for dataset protection in speech and broader biometric verification, facilitating more responsible dataset sharing and trading.

\section*{Acknowledgments}
We are sincerely grateful to Dr. Ziqi Zhang, Prof. Yong Jiang (Tsinghua University), Prof. Baoyuan Wu (The Chinese University of Hong Kong, Shenzhen), and Prof. Zhan Qin (Zhejiang University) for their valuable comments and suggestions on an early draft of this paper.

{\small \bibliographystyle{IEEEtran}}
\bibliography{main}

\newpage
\appendices
\setcounter{page}{1}
\renewcommand{\thepage}{A\arabic{page}}
\setcounter{equation}{0}
\setcounter{theorem}{2}
\setcounter{remark}{3}

\numberwithin{equation}{section}

\newenvironment{restate}[3][]{\par\medskip\noindent\textbf{#2~\ref{#3} \if\relax\detokenize{#1}\relax(Restated).\else(#1, Restated).\fi}\ \itshape\ignorespaces}{\par\medskip}


\section{Proofs of the Theoretical Results}
\label{app:proofs}

This appendix collects the proofs of all theoretical results in Section~\ref{sec:theory}. For self-containedness, each result from the main text is restated before its proof, with equation numbers reproduced from the main text; the trial-complexity result (Theorem~\ref{thm:complexity}) is stated and proved here directly. Throughout, the null hypotheses $H_0$ of Proposition~\ref{prop:similarity}\&~\ref{prop:decision} are $\mathbb{E}[S_w] = \tau \cdot \mathbb{E}[S_b]$ and $\mathbb{P}(D_w = 1) = \mathbb{P}(D_b = 1)$, respectively. We first restate the sampling protocol on which most of the proofs rely.

\begin{restate}[Independent Replication]{Assumption}{assump:rep}
In each of the $m$ verification trials, the dataset owner independently re-samples the $N$ enrolled speakers and the $K$ independent non-enrolled speakers from the speaker population and re-runs the enrollment procedure, while the trigger set $\{\bm{t}_k\}_{k=1}^{K}$ is kept fixed across trials. Accordingly, the per-trial observations are independent and identically distributed across trials, whereas the watermark-related and the independent quantities obtained within the same trial are allowed to be arbitrarily dependent.
\end{restate}

\begin{restate}[A Sufficient Condition for Coverage]{Proposition}{prop:coverage}
Suppose that there exist a metric $\mathsf{d}(\cdot,\cdot)$ on the embedding space of the suspicious model $f$ and a strictly decreasing function $\psi$ such that $\texttt{sim}(\bm{u}, \bm{v}) = \psi(\mathsf{d}(\bm{u}, \bm{v}))$ for all embeddings $\bm{u}$ and $\bm{v}$ (\eg, the cosine similarity on $\ell_2$-normalized embeddings with the angular metric), and let $R \triangleq \psi^{-1}(T)$ denote the acceptance radius induced by the decision threshold $T$. Let $\mathcal{A} \triangleq \{\bm{c}_j\}_{j=1}^{K}$ be an arbitrary collection of anchor points in this space, let $\delta \triangleq \max_{j} \mathsf{d}(f(\bm{t}_j), \bm{c}_j)$ measure how closely $f$ places each trigger to its anchor, let $\bm{V}$ denote the voiceprint of a randomly enrolled speaker, and define the covering function $\mathrm{Cov}(r) \triangleq \mathbb{P}(\min_{\bm{c} \in \mathcal{A}} \mathsf{d}(\bm{V}, \bm{c}) \leq r)$. Then the following statements hold. \textbf{(i)} If $R > \delta$, the probability that at least one trigger is accepted by the enrolled speaker satisfies $q_w \triangleq \mathbb{P}(\exists j: \texttt{sim}(f(\bm{t}_j), \bm{V}) \geq T) \geq \mathrm{Cov}(R - \delta)$. \textbf{(ii)} If the $N$ enrolled voiceprints are independent and identically distributed copies of $\bm{V}$, the probability that the trigger sequence is accepted in the 1-to-$N$ scenario is at least $1 - \left(1 - \mathrm{Cov}(R - \delta)\right)^{N}$. \textbf{(iii)} Let $\mathcal{A}'$ be another collection of anchor points with $\mathcal{A} \subseteq \mathcal{A}'$, and let $\mathrm{Cov}'$ denote its covering function defined in the same way. Then we have $\mathrm{Cov}'(r) \geq \mathrm{Cov}(r)$ for every $r \geq 0$.
\end{restate}

\begin{proof}
\textbf{(i)} On the event $\{\min_{\bm{c} \in \mathcal{A}} \mathsf{d}(\bm{V}, \bm{c}) \leq R - \delta\}$, there exists an index $j^\star$ with $\mathsf{d}(\bm{V}, \bm{c}_{j^\star}) \leq R - \delta$. By the definition of $\delta$ and the triangle inequality, $\mathsf{d}\big(f(\bm{t}_{j^\star}), \bm{V}\big) \leq \mathsf{d}\big(f(\bm{t}_{j^\star}), \bm{c}_{j^\star}\big) + \mathsf{d}\big(\bm{c}_{j^\star}, \bm{V}\big) \leq \delta + (R - \delta) = R$. Since $\texttt{sim} = \psi \circ \mathsf{d}$ with $\psi$ strictly decreasing, this implies $\texttt{sim}\big(f(\bm{t}_{j^\star}), \bm{V}\big) \geq \psi(R) = T$, so at least one trigger is accepted. Taking probabilities yields $q_w \geq \mathrm{Cov}(R - \delta)$.
 
\textbf{(ii)} For each enrolled speaker $i$, let $A_i$ denote the event that some trigger is accepted by the voiceprint $\bm{V}_i$. Given the fixed model $f$ and the fixed trigger set, $A_i$ is a deterministic function of $\bm{V}_i$ alone, so the independence of $\{\bm{V}_i\}_{i=1}^{N}$ implies the independence of $\{A_i\}_{i=1}^{N}$, each occurring with probability $q_w$. The trigger sequence is accepted in the 1-to-$N$ scenario if at least one $A_i$ occurs, whose probability equals $1 - (1 - q_w)^{N} \geq 1 - \big(1 - \mathrm{Cov}(R-\delta)\big)^{N}$ by statement (i) and the monotonicity of $x \mapsto 1-(1-x)^N$ on $[0,1]$.
 
\textbf{(iii)} Since $\mathcal{A} \subseteq \mathcal{A}'$, we have $\min_{\bm{c} \in \mathcal{A}'} \mathsf{d}(\bm{V}, \bm{c}) \leq \min_{\bm{c} \in \mathcal{A}} \mathsf{d}(\bm{V}, \bm{c})$ pointwise, so the event $\{\min_{\bm{c} \in \mathcal{A}} \mathsf{d}(\bm{V}, \bm{c}) \leq r\}$ is contained in $\{\min_{\bm{c} \in \mathcal{A}'} \mathsf{d}(\bm{V}, \bm{c}) \leq r\}$, and the claim follows from the monotonicity of probability measures.
\end{proof}

\begin{restate}[Similarity-available Verification]{Theorem}{thm:similarity}
Considering a suspicious model $f$ with the similarity function \texttt{sim} in the 1-to-$N$ speaker verification scenario, let $S_w^{(i)}$ and $S_b^{(i)}$ denote the similarities of the trigger sequence and of the independent speaker sequence in the $i$-th trial, as defined in Proposition~\ref{prop:similarity}, let $D^{(i)} \triangleq \mathbb{I}\{S_w^{(i)} > \tau \cdot S_b^{(i)}\}$ denote the win event at the $\tau$-certainty level, and let the watermark success rate $W \triangleq \frac{1}{m} \sum_{i=1}^{m} D^{(i)}$ denote the empirical estimate of $p_\tau \triangleq \mathbb{P}(S_w > \tau \cdot S_b)$ over $m$ trials. Consider the hypotheses $H_0'\colon p_\tau = \tfrac{1}{2}$ against $H_1'\colon p_\tau > \tfrac{1}{2}$, the counterparts of the hypotheses in Proposition~\ref{prop:similarity} at the resolution of win events. Suppose that Assumption~\ref{assump:rep} holds and that $m$ is sufficiently large for the normal approximation of a binomial proportion to be adequate. Then the dataset owner can reject $H_0'$ at the significance level $\alpha$ if and only if
\begin{equation}
W > \frac{1}{2} + \frac{t_{1-\alpha}}{2 \sqrt{m - 1 + t_{1-\alpha}^{2}}},
\tag{\ref{eq.thm1}}
\end{equation}
where $t_{1-\alpha}$ is the ($1-\alpha$)-quantile of the $t$-distribution with $(m-1)$ degrees of freedom, adopted as a conventional studentized calibration of the binomial proportion.
\end{restate}

\begin{proof}
Throughout the proof we write $t \triangleq t_{1-\alpha}$, and we assume $\alpha < 1/2$ so that $t > 0$. Under Assumption~\ref{assump:rep}, the per-trial win indicators $D^{(i)} \triangleq \mathbb{I}\{S_w^{(i)} > \tau \cdot S_b^{(i)}\}$, $i = 1, \cdots, m$, are independent and identically distributed Bernoulli variables with mean $p_\tau$, and $W = \frac{1}{m}\sum_{i=1}^{m} D^{(i)}$. Since $(D^{(i)})^2 = D^{(i)}$, the sample variance satisfies
\begin{equation}
s^2 = \frac{1}{m-1}\sum_{i=1}^{m}\big(D^{(i)} - W\big)^2 = \frac{m}{m-1}\, W(1-W). \nonumber
\end{equation}
The studentized test rejects $H_0'$ when $\sqrt{m}\,\big(W - \tfrac{1}{2}\big)\big/ s > t$. We first dispose of the degenerate cases. If $W = 0$, then $s = 0$ and $W < \tfrac{1}{2}$; the one-sided statistic is taken as $-\infty$, the test does not reject, and Eq.~(\ref{eq.thm1}) also fails, so the equivalence holds. If $W = 1$, then $s = 0$ and the statistic diverges to $+\infty$, so the test rejects; meanwhile $t\big/\big(2\sqrt{m-1+t^2}\big) < \tfrac{1}{2}$ for every $m \geq 2$, so $W = 1$ satisfies Eq.~(\ref{eq.thm1}) and the equivalence holds.

Moreover, for $W \in (0, 1)$ with $W \leq \tfrac{1}{2}$, the statistic is nonpositive and the test does not reject, while Eq.~(\ref{eq.thm1}) fails. It therefore suffices to consider $W \in \big(\tfrac{1}{2}, 1\big)$, where $s > 0$ and rejection is equivalent to
\begin{align}
m\big(W - \tfrac{1}{2}\big)^2 &> t^2 s^2 = \frac{t^2 m}{m-1}\, W(1-W) \nonumber \\
\Longleftrightarrow \; (m-1)\big(W - \tfrac{1}{2}\big)^2 &> t^2\, W(1-W). \nonumber
\end{align}
Writing $x \triangleq W - \tfrac{1}{2} \in \big(0, \tfrac{1}{2}\big)$ and using $W(1-W) = \tfrac{1}{4} - x^2$, the inequality becomes $(m-1)\, x^2 > t^2 \big(\tfrac{1}{4} - x^2\big)$, \ie, $\big(m-1+t^2\big)\, x^2 > \tfrac{t^2}{4}$. Since $x > 0$, this holds if and only if $x > t\big/\big(2\sqrt{m-1+t^2}\big)$, which is Eq.~(\ref{eq.thm1}). Combining the cases completes the proof.
\end{proof}

\vspace{0.3em}
\noindent \textbf{A Sufficient Condition for the Benign Regime.} If, conditionally on each realized enrollment, the similarity of the trigger sequence does not exceed $\tau$ times the median similarity of an independent speaker sequence, then the win event $\{S_w > \tau \cdot S_b\}$ requires $S_b$ to fall strictly below a point no larger than its conditional median, whence $\mathbb{P}(S_w > \tau \cdot S_b \mid \text{enrollment}) \leq \tfrac{1}{2}$; taking expectation over enrollments gives $p_\tau \leq \tfrac{1}{2}$, so $H_0'$ is the least-favorable boundary of this regime and the test in Theorem~\ref{thm:similarity} is conservative for it. The two benign scenarios in our evaluation (independent triggers and independent models) corroborate the regime empirically.

\begin{restate}[Decision-only Verification]{Theorem}{thm:decision}
Considering the 1-to-$N$ speaker verification scenario in Proposition~\ref{prop:decision}, let $\{(D_b^{(i)}, D_w^{(i)})\}_{i=1}^{m}$ denote the paired decisions collected in $m$ verification trials. Let $W \triangleq \frac{1}{m}\sum_{i=1}^{m} D_w^{(i)}$, $F \triangleq \frac{1}{m}\sum_{i=1}^{m} D_b^{(i)}$, and $\rho \triangleq \frac{1}{m}\sum_{i=1}^{m} D_w^{(i)} D_b^{(i)}$ denote the watermark success rate, the independent acceptance rate, and the joint acceptance rate, respectively. Assuming that Assumption~\ref{assump:rep} holds and that at least one discordant trial occurs, \ie, $W + F > 2\rho$, the following three statements hold.
\begin{enumerate}
\item[(i)] The one-tailed Wilcoxon signed-rank statistic adopted in Section~\ref{sec:verification}, computed with zero differences discarded and with the standard correction for tied absolute differences, admits the closed form $z = \sqrt{m}\,(W - F) \big/ \sqrt{W + F - 2\rho}$.
\item[(ii)] The dataset owner can reject the null hypothesis $H_0$ in Proposition~\ref{prop:decision} at the significance level $\alpha$ if and only if the watermark success rate $W$ satisfies
\begin{equation}\tag{\ref{eq.thm2}}
W > F + \frac{z_{1-\alpha}^{2} + \sqrt{z_{1-\alpha}^{4} + 8 m z_{1-\alpha}^{2} (F - \rho)}}{2m},
\end{equation}
where $z_{1-\alpha}$ is the ($1-\alpha$)-quantile of the standard normal distribution.
\item[(iii)] In particular, whenever every accepted independent sequence is accompanied by an accepted trigger sequence, \ie, $F = \rho$, the condition in (ii) reduces to $W > F + z_{1-\alpha}^{2} / m$, and further to $W > z_{1-\alpha}^{2} / m$ when no independent speaker is accepted in any trial.
\end{enumerate}
\end{restate}

\begin{proof}
Write $n_{10} \triangleq \#\{i: D_w^{(i)} = 1, D_b^{(i)} = 0\}$, $n_{01} \triangleq \#\{i: D_w^{(i)} = 0, D_b^{(i)} = 1\}$, and $n_d \triangleq n_{10} + n_{01}$. Since $D_w^{(i)}\big(1 - D_b^{(i)}\big) = D_w^{(i)} - D_w^{(i)}D_b^{(i)}$, summing over $i$ gives $n_{10} = m(W - \rho)$ and, symmetrically, $n_{01} = m(F - \rho)$; hence $n_d = m(W + F - 2\rho)$, which is a positive integer under the stated condition. We also record that $\rho \leq \min(W, F)$, since $D_w^{(i)}D_b^{(i)} \leq D_b^{(i)}$ and $D_w^{(i)}D_b^{(i)} \leq D_w^{(i)}$ pointwise; in particular $F - \rho \geq 0$.

\vspace{0.3em}
\noindent \textbf{Statement (i).} The paired differences $D_w^{(i)} - D_b^{(i)}$ take values in $\{-1, 0, 1\}$. Discarding the zero differences leaves $n_d$ observations whose absolute values all equal one and are therefore mutually tied; under mid-rank assignment, each receives the average rank $\frac{1}{n_d}\sum_{r=1}^{n_d} r = \frac{n_d+1}{2}$. The one-tailed signed-rank statistic is the sum of the ranks of the positive differences, $W^{+} = n_{10} \cdot \frac{n_d+1}{2}$. Its null mean is $\mathbb{E}_{H_0}[W^{+} \mid n_d] = \frac{n_d(n_d+1)}{4}$, and the tie-corrected null variance is
\begin{align}
\sigma^2 &= \frac{n_d(n_d+1)(2n_d+1)}{24} - \frac{n_d^3 - n_d}{48} \nonumber\\
&= \frac{2n_d(n_d+1)(2n_d+1) - (n_d^3 - n_d)}{48} \nonumber\\
&= \frac{3n_d(n_d+1)^2}{48} = \frac{n_d(n_d+1)^2}{16},
\end{align}
where the tie-correction term $\frac{\sum_g (t_g^3 - t_g)}{48}$ is evaluated at the single tie group of size $t_g = n_d$. Standardizing,
\begin{align}
z &= \frac{W^{+} - \frac{n_d(n_d+1)}{4}}{\sqrt{\sigma^2}} = \frac{\frac{n_d+1}{4}\big(2n_{10} - n_d\big)}{\frac{(n_d+1)\sqrt{n_d}}{4}} \nonumber\\
&= \frac{2n_{10} - n_d}{\sqrt{n_d}}.
\end{align}
Substituting $n_{10} = m(W-\rho)$ and $n_d = m(W + F - 2\rho)$ gives $2n_{10} - n_d = m(W - F)$ and hence $z = \sqrt{m}\,(W - F)/\sqrt{W + F - 2\rho}$, as claimed.

\vspace{0.3em}
\noindent \textbf{Statement (ii).} The test rejects when $z > z_{1-\alpha}$, and we assume $\alpha < 1/2$ so that $z_{1-\alpha} > 0$; rejection therefore requires $W > F$. Write $u \triangleq W - F > 0$ and $\varphi \triangleq F - \rho \geq 0$, so that $W + F - 2\rho = u + 2\varphi$. On $\{u > 0\}$, squaring $\sqrt{m}\,u > z_{1-\alpha}\sqrt{u + 2\varphi}$ is an equivalence and yields $m u^2 - z_{1-\alpha}^2 u - 2 z_{1-\alpha}^2 \varphi > 0$. The roots of this quadratic in $u$ are
\begin{equation}
u_{1,2} = \frac{z_{1-\alpha}^{2} \pm \sqrt{z_{1-\alpha}^{4} + 8 m z_{1-\alpha}^{2}\varphi}}{2m},
\end{equation}
and since $\varphi \geq 0$ implies $\sqrt{z_{1-\alpha}^{4} + 8 m z_{1-\alpha}^{2}\varphi} \geq z_{1-\alpha}^{2}$, we have $u_1 \leq 0 < u_2$. The quadratic is positive precisely for $u < u_1$ or $u > u_2$, and the constraint $u > 0$ excludes the former branch, so rejection occurs if and only if $u > u_2$, which is exactly Eq.~(\ref{eq.thm2}). Conversely, $u > u_2$ implies $u > 0$ and the squared inequality, hence $z > z_{1-\alpha}$.
 
\vspace{0.3em}
\noindent \textbf{Statement (iii).} If $F = \rho$, then $\varphi = 0$ and the square root reduces to $z_{1-\alpha}^{2}$, so the margin becomes $\frac{2z_{1-\alpha}^2}{2m} = \frac{z_{1-\alpha}^2}{m}$ and the condition reads $W > F + z_{1-\alpha}^{2}/m$. If in addition $F = 0$, then $0 \leq \rho \leq F = 0$ forces $\rho = 0$, and the condition further reduces to $W > z_{1-\alpha}^{2}/m$.
\end{proof}

\begin{restate}[Exact Certificate]{Corollary}{cor:exact}
Under Assumption~\ref{assump:rep}, let $n_{10}$ and $n_{01}$ denote the numbers of trials with $(D_w^{(i)}, D_b^{(i)})$ equal to $(1,0)$ and $(0,1)$ respectively, and let $n_d \triangleq n_{10} + n_{01}$. Consider the exact conditional test that rejects $H_0$ when $n_{10} \geq b_{\alpha}(n_d)$, where $b_{\alpha}(n) \triangleq \min\{j \in \{0,\ldots,n\} : 2^{-n}\sum_{l=j}^{n}\binom{n}{l} \leq \alpha\}$ and $\min\emptyset \triangleq +\infty$. Then this test satisfies $\mathbb{P}_{H_0}(\text{reject}) \leq \alpha$ for every $m \geq 1$, without invoking any normal approximation. In particular, when $F = 0$, it rejects $H_0$ at the significance level $\alpha$ if and only if
\begin{equation}\tag{\ref{eq.cor}}
W \geq \frac{1}{m}\log_2\frac{1}{\alpha}.
\end{equation}
\end{restate}

\begin{proof}
Under Assumption~\ref{assump:rep}, the $m$ trials are independent and identically distributed, and each trial falls into the category $(1,0)$ with probability $\pi_{10}$, into $(0,1)$ with probability $\pi_{01}$, and into a concordant category otherwise. By the standard conditioning property of independent categorical trials, conditioned on $n_d = n$ the count $n_{10}$ follows the binomial distribution $B\big(n, \frac{\pi_{10}}{\pi_{10}+\pi_{01}}\big)$. Under $H_0$, which by the marginal-homogeneity identity $\mathbb{P}(D_w{=}1) - \mathbb{P}(D_b{=}1) = \pi_{10} - \pi_{01}$ equals the condition $\pi_{10} = \pi_{01}$, this conditional distribution is $B(n, 1/2)$.
 
Let $B_n \sim B(n, 1/2)$ and $G_n(j) \triangleq \mathbb{P}(B_n \geq j) = 2^{-n}\sum_{l=j}^{n}\binom{n}{l}$, which is strictly decreasing in $j$ on $\{0, 1, \ldots, n\}$. By the definition of $b_{\alpha}(n)$ and this strict monotonicity, $\{j : G_n(j) \leq \alpha\} = \{j : j \geq b_{\alpha}(n)\}$. Conditioned on $n_d = n$, the rejection event $\{n_{10} \geq b_{\alpha}(n)\}$ therefore has probability $G_n\big(b_{\alpha}(n)\big) \leq \alpha$, where the inequality holds by the definition of $b_{\alpha}(n)$ and trivially when $b_{\alpha}(n) = +\infty$. Taking the expectation over the distribution of $n_d$ yields $\mathbb{P}_{H_0}(\text{reject}) = \mathbb{E}\big[\mathbb{P}_{H_0}(\text{reject} \mid n_d)\big] \leq \alpha$, and no normal approximation has been used at any step.
 
For the special case, if $F = 0$ then $\sum_{i=1}^{m} D_b^{(i)} = 0$, which forces $n_{01} = 0$ and $\rho = 0$, hence $n_{10} = \sum_{i=1}^{m} D_w^{(i)} = mW$ and $n_d = mW$. The exact conditional p-value is then $G_{n_d}(n_d) = 2^{-n_d} = 2^{-mW}$, so rejection at the significance level $\alpha$ occurs if and only if $2^{-mW} \leq \alpha$, that is, if and only if $mW \geq \log_2(1/\alpha)$.
\end{proof}

For the next proposition, we also restate the simplifying assumption on which it relies; it is used only here.

\begin{restate}[Conditional Independence]{Assumption}{assump:ci}
Given the suspicious model $f$, the events that the trigger sequence is accepted by distinct enrolled speakers are mutually independent with a common probability $q_w \in (0,1)$, the events that the independent speaker sequence is accepted by distinct enrolled speakers are mutually independent with a common probability $q_b \in (0,1)$, and $D_w$ is independent of $D_b$.
\end{restate}

\begin{restate}[Effect of the Enrollment Size]{Proposition}{prop:enrollment}
Assuming that Assumption~\ref{assump:ci} holds and that the watermark is effective in the sense that $q_w > q_b$, the discordance odds ratio $\theta(N) \triangleq \pi_{10}(N) / \pi_{01}(N)$ is strictly increasing in $N$, where $\pi_{10}(N) \triangleq \mathbb{P}(D_w = 1, D_b = 0)$ and $\pi_{01}(N) \triangleq \mathbb{P}(D_w = 0, D_b = 1)$. Consequently, conditioned on any number of discordant trials for which the rejection region is non-degenerate, the power of the test in Theorem~\ref{thm:decision} increases with $N$.
\end{restate}

\begin{proof}
Under Assumption~\ref{assump:ci}, the trigger sequence fails in a trial if and only if it is rejected by all $N$ enrolled speakers, so $\mathbb{P}(D_w = 0) = (1 - q_w)^{N}$, and likewise $\mathbb{P}(D_b = 0) = (1 - q_b)^{N}$. Since $D_w$ and $D_b$ are independent, $\pi_{10}(N) = \big[1 - (1 - q_w)^{N}\big](1 - q_b)^{N}$ and $\pi_{01}(N) = (1 - q_w)^{N}\big[1 - (1 - q_b)^{N}\big]$. Letting $a \triangleq -\ln(1 - q_w)$ and $\nu \triangleq -\ln(1 - q_b)$, both positive, we obtain
\begin{equation}
\theta(N) = \frac{1 - (1-q_w)^N}{(1-q_w)^N} \cdot \frac{(1-q_b)^N}{1 - (1-q_b)^N} = \frac{e^{aN} - 1}{e^{\nu N} - 1}.
\end{equation}
 
Treating $N$ as a positive real number and differentiating the logarithm of $\theta$,
\begin{align}
\frac{\partial}{\partial N}\ln\theta(N) &= \frac{a e^{aN}}{e^{aN} - 1} - \frac{\nu e^{\nu N}}{e^{\nu N} - 1} \nonumber\\
&= \frac{1}{N}\Big[h(aN) - h(\nu N)\Big], h(s) \triangleq \frac{s}{1 - e^{-s}},
\end{align}
where we used $\frac{s e^{s}}{e^{s}-1} = \frac{s}{1 - e^{-s}}$. To show that $h$ is strictly increasing on $(0,\infty)$, compute $h'(s) = \chi(s) / (1 - e^{-s})^{2}$ with $\chi(s) \triangleq 1 - (1 + s)e^{-s}$; since $\chi(0) = 0$ and $\chi'(s) = s e^{-s} > 0$ for every $s > 0$, we have $\chi(s) > 0$ and hence $h'(s) > 0$. Because $q_w > q_b$ implies $a > \nu$, it follows that $h(aN) > h(\nu N)$ and therefore that $\theta(N)$ is strictly increasing in $N$.
 
For the power claim, the conditioning argument in the proof of Corollary~\ref{cor:exact} applies verbatim under the alternative and gives $n_{10} \mid n_d \sim B\big(n_d, \theta(N)/(1 + \theta(N))\big)$. The rejection condition $z > z_{1-\alpha}$ of Theorem~\ref{thm:decision} is, by statement (i), equivalent to $2n_{10} - n_d > z_{1-\alpha}\sqrt{n_d}$, \ie, to $n_{10} > \frac{1}{2}\big(n_d + z_{1-\alpha}\sqrt{n_d}\big)$, which defines an upper set in $n_{10}$. Since $\theta \mapsto \theta/(1+\theta)$ is strictly increasing on $(0,\infty)$, and since for $B \sim B(n, q)$ and $1 \leq j \leq n$ we have $\frac{\partial}{\partial q}\mathbb{P}(B \geq j) = n\binom{n-1}{j-1}q^{j-1}(1-q)^{n-j} > 0$, the conditional power is strictly increasing in $\theta(N)$, and therefore increasing in $N$.
\end{proof}

The last result of this appendix addresses how many trials suffice for the audit to succeed with a prescribed probability, as previewed in Section~\ref{sec:theory}. It applies to the exact test of Corollary~\ref{cor:exact} and again involves no normal approximation.

\begin{theorem}[Trial Complexity of the Decision-only Audit]
\label{thm:complexity}
Suppose Assumption~\ref{assump:rep} holds, and suppose the data-generating distribution satisfies $\pi_d \triangleq \pi_{10} + \pi_{01} > 0$ and $\varepsilon \triangleq \frac{\pi_{10}}{\pi_d} - \frac{1}{2} > 0$, where $\pi_{10} \triangleq \mathbb{P}(D_w^{(i)} = 1, D_b^{(i)} = 0)$ and $\pi_{01} \triangleq \mathbb{P}(D_w^{(i)} = 0, D_b^{(i)} = 1)$ denote the per-trial discordance probabilities. Fix a significance level $\alpha \in (0,1)$ and a target failure probability $\tilde\beta \in (0,1)$. If the number of trials satisfies
\begin{equation}\label{eq.complexity}
m \;\geq\; \max\left\{ \frac{2}{\pi_d^{2}} \ln\frac{2}{\tilde\beta}, \;\; \frac{2}{\pi_d \, \varepsilon^{2}} \left( \sqrt{\tfrac{1}{2}\ln\tfrac{1}{\alpha}} + \sqrt{\tfrac{1}{2}\ln\tfrac{2}{\tilde\beta}} \right)^{2} \right\},
\end{equation}
then the exact conditional test of Corollary~\ref{cor:exact} rejects $H_0$ with probability at least $1 - \tilde\beta$.
\end{theorem}

\begin{proof}
Write $c_{\alpha} \triangleq \sqrt{\tfrac{1}{2}\ln\tfrac{1}{\alpha}}$, $c_{\beta} \triangleq \sqrt{\tfrac{1}{2}\ln\tfrac{2}{\tilde\beta}}$, $n_0 \triangleq (c_{\alpha} + c_{\beta})^2 / \varepsilon^2$, and $n^\star \triangleq \lceil m\pi_d/2 \rceil$. The second term of Eq.~(\ref{eq.complexity}) guarantees $m\pi_d/2 \geq n_0$, hence $n^\star \geq n_0$. The proof proceeds in three steps.

\vspace{0.3em} 
\noindent \textbf{Step 1 (an upper bound on the critical value).} Let $B_n \sim B(n, 1/2)$. Hoeffding's inequality gives $\mathbb{P}\big(B_n \geq \tfrac{n}{2} + t\big) \leq \exp(-2t^{2}/n)$ for every $t \geq 0$. Taking $t = \sqrt{n}\,c_{\alpha}$ yields $\mathbb{P}\big(B_n \geq \tfrac{n}{2} + \sqrt{n}\,c_{\alpha}\big) \leq \alpha$, so by the definition of $b_{\alpha}(n)$ as the smallest integer threshold achieving conditional level $\alpha$, we have $b_{\alpha}(n) \leq \big\lceil \tfrac{n}{2} + \sqrt{n}\,c_{\alpha} \big\rceil$, and therefore $b_{\alpha}(n) - 1 \leq \tfrac{n}{2} + \sqrt{n}\,c_{\alpha}$.

\begin{table*}[!t]
\caption{The WSR (\%) and EER (\%) of our CBW with respect to different trigger patterns on the TIMIT dataset.}
\vspace{-0.5em}
 \centering
\scalebox{0.9}{
\begin{tabular}{c|c|c|ccc}
\toprule
Model$\downarrow$                       & Verification Scenario$\downarrow$  & Metric$\downarrow$, Trigger Pattern$\rightarrow$ & One-hot-spectrum Noise & Multi-hot-spectrum Noise & Gaussian Noise \\ \cline{1-6}
\multirow{4}{*}{LSTM}        & \multirow{2}{*}{1-to-1} & EER (\%)                        & 6.4                 & 6.1               & 5.9         \\
                             &                         & WSR (\%)                        & 80.7                  & 69.3                & 71.3          \\ \cline{2-6} 
                             & \multirow{2}{*}{1-to-5} & EER (\%)                        & 5.2                 & 4.8               & 5.0         \\
                             &                         & WSR (\%)                        & 100.0                      & 100.0                    & 100.0              \\ \hline
\multirow{4}{*}{Ecapa-tdnn} & \multirow{2}{*}{1-to-1} & EER (\%)                        & 5.5                 & 5.4               & 5.5         \\
                             &                         & WSR (\%)                        & 58.7                  & 46.7                & 44.0           \\ \cline{2-6} 
                             & \multirow{2}{*}{1-to-5} & EER (\%)                        & 4.5                 & 4.3               & 4.6         \\
                             &                         & WSR (\%)                        & 100.0                      & 88.3                & 87.6          \\ \hline
\multirow{4}{*}{CAM++}       & \multirow{2}{*}{1-to-1} & EER (\%)                        & 7.1                 & 9.3               & 7.6         \\
                             &                         & WSR (\%)                        & 55.3                  &41.7                & 48.0           \\ \cline{2-6} 
                             & \multirow{2}{*}{1-to-5} & EER (\%)                        & 6.2                  & 7.1              & 6.7         \\
                             &                         & WSR (\%)                        & 98.3                  & 87.3                & 90.3          \\ 
                             \bottomrule
\end{tabular}}
\label{tab:trigger pattern}
\vspace{-0.8em}
\end{table*}

\vspace{0.3em} 
\noindent \textbf{Step 2 (conditional type II error).} By the conditioning argument in the proof of Corollary~\ref{cor:exact}, under the alternative $n_{10} \mid n_d = n \sim B(n, q^\star)$ with $q^\star = \tfrac{1}{2} + \varepsilon$. The test fails to reject exactly when $n_{10} \leq b_{\alpha}(n) - 1$, and by Step 1 this event is contained in $\big\{ n_{10} \leq \tfrac{n}{2} + \sqrt{n}\,c_{\alpha} \big\} = \big\{ n_{10} \leq n q^\star - \sqrt{n}\big(\sqrt{n}\,\varepsilon - c_{\alpha}\big) \big\}$. For $n \geq n_0$ we have $\sqrt{n}\,\varepsilon - c_{\alpha} \geq c_{\beta} \geq 0$, so Hoeffding's inequality for the lower tail of $B(n, q^\star)$ yields
\begin{align}
\mathbb{P}\big( n_{10} < b_{\alpha}(n) \,\big|\, n_d = n \big) \;&\leq\; \exp\!\Big( -2\big(\sqrt{n}\,\varepsilon - c_{\alpha}\big)^{2} \Big) \nonumber\\
&\leq\; \exp\!\big( -2c_{\beta}^{2} \big) \;=\; \frac{\tilde\beta}{2}.
\end{align}
Since $n \mapsto \sqrt{n}\,\varepsilon - c_{\alpha}$ is increasing, this bound holds uniformly for all $n \geq n^\star \geq n_0$.
 
\vspace{0.3em} 
\noindent \textbf{Step 3 (enough discordant trials).} Under Assumption~\ref{assump:rep}, $n_d \sim B(m, \pi_d)$. Hoeffding's inequality for the lower tail with deviation $m\pi_d/2$ gives $\mathbb{P}\big( n_d < m\pi_d/2 \big) \leq \exp\big( -2m(\pi_d/2)^{2} \big) = \exp\big( -m\pi_d^{2}/2 \big) \leq \tilde\beta/2$, where the last inequality uses the first term of Eq.~(\ref{eq.complexity}).
 
Combining the three steps, the probability that the test fails to reject is at most
\begin{align}
&\mathbb{P}\big(n_d < n^\star\big) + \sup_{n \geq n^\star} \mathbb{P}\big( n_{10} < b_{\alpha}(n) \,\big|\, n_d = n \big) \nonumber\\
&\quad\leq\; \frac{\tilde\beta}{2} + \frac{\tilde\beta}{2} \;=\; \tilde\beta,
\end{align}
which completes the proof.
\end{proof}

\begin{remark}
Two remarks are in order. First, the constants in Eq.~(\ref{eq.complexity}) are not optimized; the bound is meant to expose how the required number of trials scales with the significance level, the target failure probability, the discordance probability $\pi_d$, and the sign bias $\varepsilon$, rather than to be numerically tight. For instance, when $\pi_{01} = 0$, the exact analysis of Corollary~\ref{cor:exact} already succeeds with five accepted trials at $\alpha = 0.05$. Second, under Assumption~\ref{assump:ci} the sign bias $\varepsilon = (\theta(N) - 1) \big/ \left(2(\theta(N) + 1)\right)$ is strictly increasing in the odds ratio $\theta(N)$ and hence in $N$ by Proposition~\ref{prop:enrollment}, which lowers the second term of Eq.~(\ref{eq.complexity}); at the same time, a very large enrollment drives both acceptance rates toward one and the discordance probability $\pi_d$ toward zero, which inflates the first term. This suggests a saturation regime in which further enlarging the enrollment no longer helps the audit.
\end{remark}

\section{Ablation Studies}
\label{app:ablation}
We hereby analyze the effects of the core modules and hyper-parameters involved in our CBW-based dataset ownership verification. Unless otherwise specified, all other settings are the same as those in Section~\ref{settings}, and all experiments in this part are conducted on TIMIT due to limited space.

\begin{table}[!t]
\caption{The WSR (\%) and EER (\%) of our CBW with different cluster methods on the TIMIT dataset.}
\vspace{-0.5em}
\centering
\scalebox{1}{
\begin{tabular}{c|cc|cc|cc}
\toprule
Model$\rightarrow$                            & \multicolumn{2}{c|}{LSTM} & \multicolumn{2}{c|}{Ecapa-tdnn} & \multicolumn{2}{c}{CAM++} \\ \hline
Clustering Method$\downarrow$                   & EER          & WSR        & EER            & WSR           & EER          & WSR        \\ \hline
k-means                    & 6.4       & 80.7      & 5.5          & 58.7         & 7.1      & 55.3      \\ 
Spectral Clustering                  & 6.6       & 78.7      & 5.4          & 56.3         & 7.2       & 53.3      \\ 
GMM & 5.9       & 76.5      & 6.4          & 56.3         & 7.5       & 55.7      \\ \bottomrule
\end{tabular}}
\label{tab:cluster methods}
\vspace{-0.8em}
\end{table}

\vspace{0.3em}
\noindent \textbf{Effects of Clustering Methods.} Recall that in our second step, the training speakers are partitioned into $K$ disjoint clusters based on their feature representations through a given clustering method. We hereby evaluate the effects of this core module with three classical clustering methods, including $k$-means~\cite{lloyd1982least} (\ie, the one used before), spectral clustering~\cite{ng2001spectral}, and the Gaussian mixture model (GMM)~\cite{ouyang2004gaussian}. As shown in Table~\ref{tab:cluster methods}, all variants achieve satisfactory performance, although there are some mild fluctuations. The mild fluctuations across these three clustering methods are consistent with our design: the watermarking stage only requires a partition that groups speakers with similar feature representations (Section~\ref{sec:watermarking}), and Proposition~\ref{prop:coverage} holds for an arbitrary collection of anchor points rather than for the output of any specific clustering algorithm. In other words, what matters is how the anchors spread over the embedding space, not which classical method produces them.

\begin{figure*}[!t]
\vspace{-1em}
    \begin{minipage}[t]{0.48\linewidth}
        \centering
	\subfloat[EER (\%) of 1-to-1 scenario]{
		\includegraphics[width=0.45\linewidth]{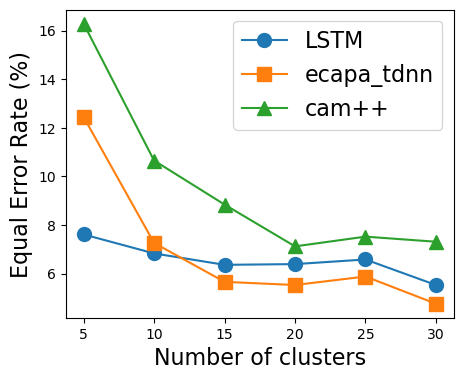}}\hspace{0.5em}
	\subfloat[WSR (\%) of 1-to-1 scenario]{
		\includegraphics[width=0.45\linewidth]{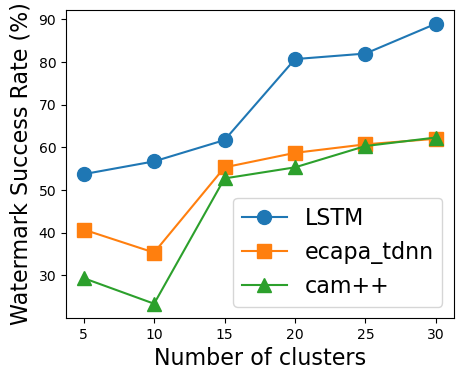}}\\
        \subfloat[EER (\%) of 1-to-5 scenario]{
		\includegraphics[width=0.45\linewidth]{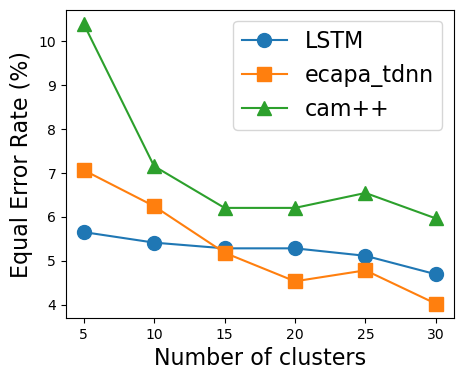}} \hspace{0.5em}
        \subfloat[WSR (\%) of 1-to-5 scenario]{
		\includegraphics[width=0.45\linewidth]{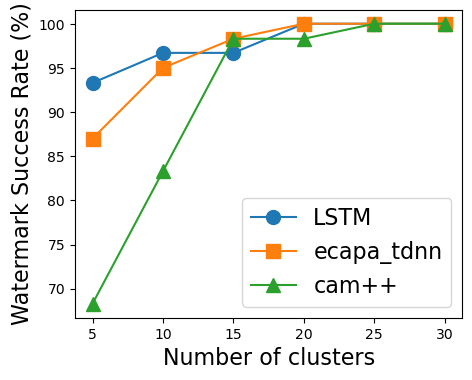}}
	\caption{The WSR (\%) and EER (\%) of our CBW with respect to different numbers of clusters (\ie, $K$) on the TIMIT dataset.}
     \label{fig:cluster_num}
    \end{minipage}\hspace{1.5em}
    \begin{minipage}[t]{0.48\linewidth}
        \centering
	\subfloat[EER (\%) of 1-to-1 scenario]{
		\includegraphics[width=0.453\linewidth]{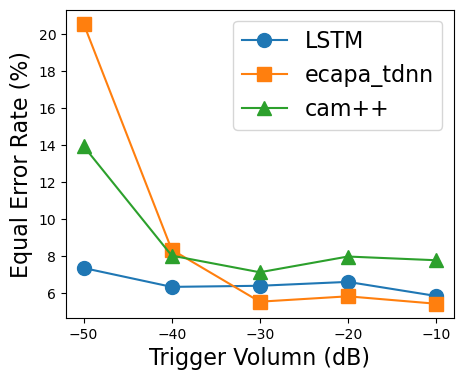}}\hspace{0.5em}
	\subfloat[WSR (\%) of 1-to-1 scenario]{
		\includegraphics[width=0.453\linewidth]{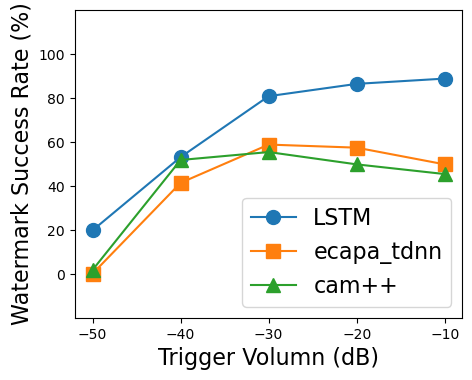}}\\
        \subfloat[EER (\%) of 1-to-5 scenario]{
		\includegraphics[width=0.453\linewidth]{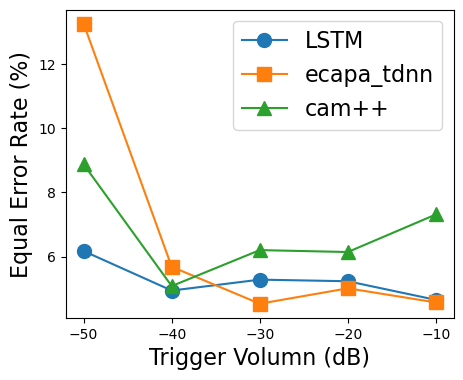}}\hspace{0.5em}
        \subfloat[WSR (\%) of 1-to-5 scenario]{
		\includegraphics[width=0.453\linewidth]{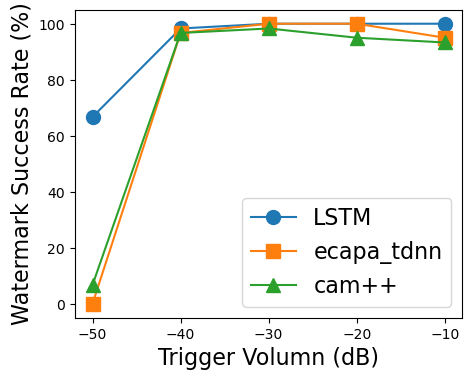}}
	\caption{The WSR (\%) and EER (\%) of our CBW with respect to different trigger volumes on the TIMIT dataset.}
 \label{fig:trigger_volume}
    \end{minipage}%
    \vspace{-0.8em}
\end{figure*}

\vspace{0.3em}
\noindent \textbf{Effects of the Number of Clusters.} We hereby explore the effects of the number of clusters $K$, varying from 5 to 30. As shown in Figure~\ref{fig:cluster_num}, the WSR generally increases and the EER generally decreases with the increase of $K$ in all evaluated model structures and verification scenarios. The WSR trend is consistent with Proposition~\ref{prop:coverage}: a finer partition places some anchor closer to any enrolled voiceprint, so the trigger set covers the embedding space better. The EER trend is also in line with the cluster-local design in Section~\ref{sec:watermarking}, since more clusters make each trigger associate with a smaller region and thus perturb the embedding geometry more locally. However, increasing $K$ also enlarges the verification overhead, since the dataset owner needs to query the suspicious model with all $K$ triggers in sequence per each verification query. There is thus a trade-off between effectiveness and overhead, and the owner can assign $K$ per specific needs in practice.

\vspace{0.3em}
\noindent \textbf{Effects of the Trigger Volume.} We hereby explore the effects of the trigger volume on our CBW. As shown in Figure~\ref{fig:trigger_volume}, the EER generally decreases and the WSR generally increases with the increase of the trigger volume. A plausible explanation is that a larger volume leads to more distinctive features that are more likely to be learned by DNNs.

\vspace{0.3em}
\noindent \textbf{Effects of the Trigger Pattern.} We hereby evaluate our CBW with different trigger patterns, including the one-hot-spectrum noise used in our main experiments, multi-hot-spectrum noise, and Gaussian noise. Multi-hot-spectrum noise is a sinusoidal signal whose spectrum contains several discrete frequency components, while Gaussian noise is a random signal whose amplitude follows a normal distribution. As shown in Table~\ref{tab:trigger pattern}, our method is effective across all three trigger patterns, although the performance varies to some extent. One possible explanation is that the multiple activation spectra of multi-hot-spectrum noise and the randomness of Gaussian noise make them relatively harder to be learned as triggers by DNNs.

\begin{figure*}[!t]
	\centering
	\subfloat[EER (\%) of 1-to-1 scenario]{
		\includegraphics[width=0.22\linewidth]{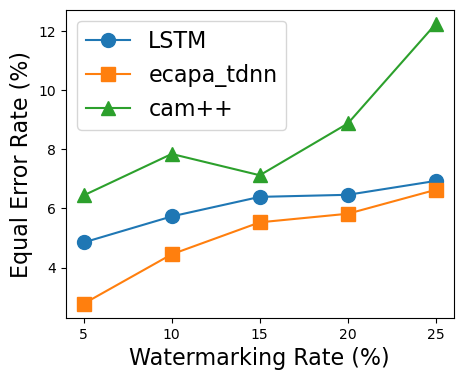}}
	\subfloat[WSR (\%) of 1-to-1 scenario]{
		\includegraphics[width=0.22\linewidth]{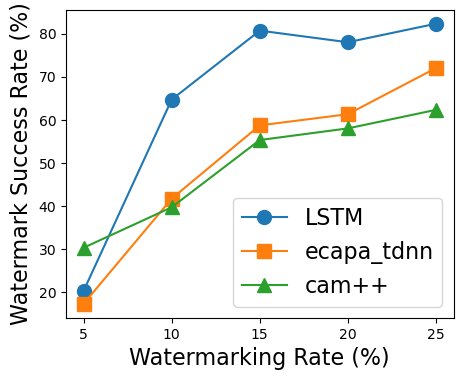}} \hspace{1em}
       \subfloat[EER (\%) of 1-to-5 scenario]{
		\includegraphics[width=0.22\linewidth]{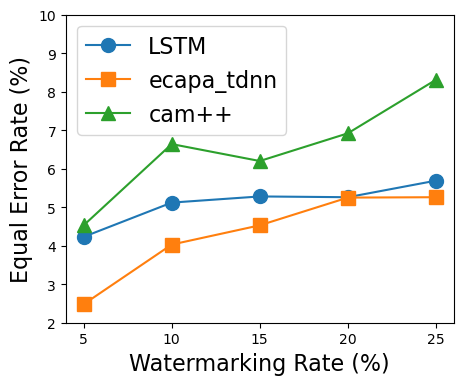}}
        \subfloat[WSR (\%) of 1-to-5 scenario]{
		\includegraphics[width=0.22\linewidth]{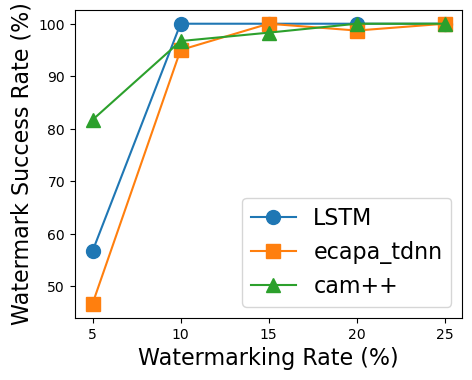}}
	\caption{The WSR (\%) and EER (\%) of our CBW with respect to different watermarking rates on the TIMIT dataset.}
 \label{fig:poison_rate}
\end{figure*}

\vspace{0.3em}
\noindent \textbf{Effects of the Watermarking Rate.} We hereby explore the effects of the watermarking rate $\gamma$, varying from 5\% to 25\%. As shown in Figure~\ref{fig:poison_rate}, the WSR increases with the increase of $\gamma$, whereas the EER also rises with it. This trade-off is the coverage-fidelity tension distilled in Section~\ref{sec:requirements} made quantitative: a larger $\gamma$ strengthens each trigger-cluster association at the cost of perturbing more of the released data.

\vspace{0.3em}
\noindent \textbf{Effects of the Number of Trials.} We hereby discuss the effects of the number of trials $m$ used in each verification. Specifically, we vary $m$ from 20 to 100 and report the p-values under both access levels. As shown in Tables~\ref{tab:T_test sample num}-\ref{tab:W_test sample num}, the p-values in the Dataset Stealing scenario shrink rapidly as $m$ grows, which is consistent with Theorem~\ref{thm:similarity} and Theorem~\ref{thm:decision}, where the required margin over the independent acceptance level shrinks as $m$ increases. However, more trials also lead to more model queries and extra overhead. Accordingly, the dataset owner can assign $m$ based on specific requirements.

\begin{table*}[!t]
\vspace{-2em}
\caption{The verification effectiveness of similarity-available dataset verification with different numbers of trials on TIMIT.}
\vspace{-0.5em}
\tabcolsep=3.5mm
 \centering
\scalebox{0.85}{
\begin{tabular}{c|c|c|ccccc}
\toprule
Model$\downarrow$                      & Verification Scenario$\downarrow$  & Scenario$\downarrow$, Number of Trials$\rightarrow$  & 20       & 40       & 60       & 80        & 100      \\ \hline
\multirow{6}{*}{LSTM}       & \multirow{3}{*}{1-to-1} & Independent Model           & 1        & 1        & 1        & 1         & 1        \\
                            &                         & Independent Trigger         & 1        & 1        & 1        & 1         & 1        \\
                            &                         & Dataset Stealing                       & $10^{-2}$ & $10^{-2}$ & $10^{-3}$ & $10^{-5}$  & $10^{-7}$ \\ \cline{2-8} 
                            & \multirow{3}{*}{1-to-5} & Independent Model               & 1        & 1        & 1        & 1         & 1        \\
                            &                         & Independent Trigger         & 1        & 1        & 1        & 1         & 1        \\
                            &                         & Dataset Stealing                       & $10^{-5}$ & $10^{-7}$ & $10^{-9}$ & $10^{-12}$  & $10^{-19}$ \\ \hline
\multirow{6}{*}{Ecapa-tdnn} & \multirow{3}{*}{1-to-1} & Independent Model           & 1        & 1        & 1        & 0.95   & 0.99  \\
                            &                         & Independent Trigger         & 1        & 1        & 1        & 1         & 1        \\
                            &                         & Dataset Stealing                       & $10^{-2}$ & $10^{-15}$ & $10^{-19}$ & $10^{-36}$  & $10^{-38}$ \\ \cline{2-8} 
                            & \multirow{3}{*}{1-to-5} & Independent Model           & 1        & 1        & 1        & 1         & 1        \\
                            &                         & Independent Trigger         & 1        & 1        & 1        & 1         & 1        \\
                            &                         & Dataset Stealing                       & $10^{-14}$ & $10^{-44}$ & $10^{-76}$ & $10^{-138}$ & 0        \\ \hline
\multirow{6}{*}{CAM++}      & \multirow{3}{*}{1-to-1} & Independent Model           & 1        & 1        & 1        & 1         & 1        \\
                            &                         & Independent Trigger         & 1        & 1        & 1        & 1         & 1        \\
                            &                         & Dataset Stealing                       & $10^{-5}$ & $10^{-17}$ & $10^{-20}$ & $10^{-38}$  & $10^{-49}$ \\ \cline{2-8} 
                            & \multirow{3}{*}{1-to-5} & Independent Model           & 1        & 1        & 1        & 1         & 1        \\
                            &                         & Independent Trigger         & 1        & 1        & 1        & 1         & 1        \\
                            &                         & Dataset Stealing                       & $10^{-10}$ & $10^{-82}$ & $10^{-97}$ & $10^{-168}$ & 0        \\ \bottomrule
\end{tabular}}
\label{tab:T_test sample num}
\end{table*}

\begin{table*}[!t]
\caption{The verification effectiveness of decision-only dataset verification with different numbers of trials on TIMIT.}
\vspace{-0.5em}
\tabcolsep=3.5mm
 \centering
\scalebox{0.85}{
\begin{tabular}{c|c|c|ccccc}
\toprule
Model$\downarrow$                     & Verification Scenario$\downarrow$  & Scenario$\downarrow$, Number of Trials$\rightarrow$ & 20       & 40       & 60      & 80      & 100     \\ \hline
\multirow{6}{*}{LSTM}       & \multirow{3}{*}{1-to-1} & Independent Model           & 0.72  & 0.92        & 0.95       & 0.96       & 1       \\
                            &                         & Independent Trigger         & 0.71  & 0.91  & 0.91 & 0.95  & 0.99 \\
                            &                         & Dataset Stealing                       & $10^{-2}$ & $10^{-3}$ & $10^{-4}$       & $10^{-7}$        & $10^{-10}$        \\ \cline{2-8} 
                            & \multirow{3}{*}{1-to-5} & Independent Model              & 0.75        & 0.89        & 0.99       & 0.99       & 1       \\
                            &                         & Independent Trigger         & 0.67  & 0.90        & 0.998       & 0.99       & 1       \\
                            &                         & Dataset Stealing                       & $10^{-2}$       & $10^{-4}$       & $10^{-7}$  & $10^{-9}$       & $10^{-13}$       \\ \hline
\multirow{6}{*}{Ecapa-tdnn} & \multirow{3}{*}{1-to-1} & Independent Model           & 0.72  & 0.91  & 0.98 & 0.98 & 0.99 \\
                            &                         & Independent Trigger         & 0.72  & 0.91  & 0.99 & 0.98 & 0.99 \\
                            &                         & Dataset Stealing                       & $10^{-6}$ & $10^{-12}$ & 0       & 0       & 0       \\ \cline{2-8} 
                            & \multirow{3}{*}{1-to-5} & Independent Model           & 1        & 1        & 1       & 1       & 1       \\
                            &                         & Independent Trigger         & 1        & 1        & 1       & 1       & 1       \\
                            &                         & Dataset Stealing                       & 0        & 0        & 0       & 0       & 0       \\ \hline
\multirow{6}{*}{CAM++}      & \multirow{3}{*}{1-to-1} & Independent Model           & 1        & 0.99  & 0.99 & 0.99 & 0.99 \\
                            &                         & Independent Trigger         & 0.95  & 0.99  & 0.99  & 0.99 & 0.99  \\
                            &                         & Dataset Stealing                       & $10^{-9}$ & 0        & 0       & 0       & 0       \\ \cline{2-8} 
                            & \multirow{3}{*}{1-to-5} & Independent Model           & 1        & 1        & 1       & 1       & 1       \\
                            &                         & Independent Trigger         & 1        & 1        & 1       & 1       & 1       \\
                            &                         & Dataset Stealing                       & $10^{-10}$ & 0        & 0       & 0       & 0       \\ 
                            \bottomrule
\end{tabular}}
\label{tab:W_test sample num}
\end{table*}

\section{Resistance to Watermark-removal Attacks}
\label{app:robustness}

We hereby discuss the resistance of our CBW against three representative watermark-removal attacks, including fine-tuning~\cite{liu2017neural}, model pruning~\cite{liu2018fine}, and data augmentation~\cite{park2019specaugment}.

\begin{figure*}[!t]
\vspace{-2em}
    \begin{minipage}[t]{0.48\linewidth}
        \centering
\subfloat[LSTM (1-to-1 scenario)]{
		\includegraphics[width=0.45\textwidth]{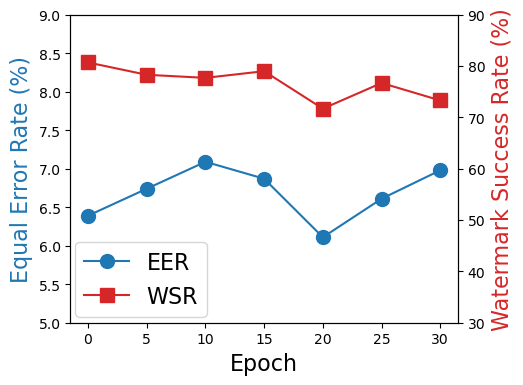}}
    \subfloat[LSTM (1-to-5 scenario)]{
		\includegraphics[width=0.45\textwidth]{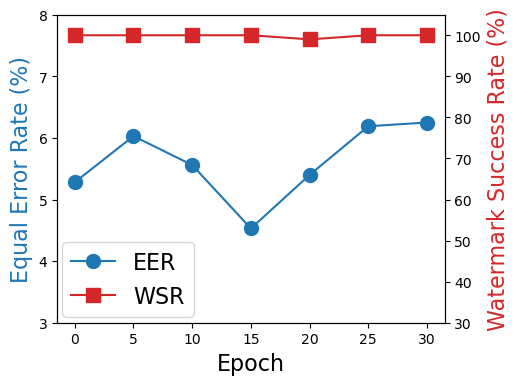}}\\	
   \subfloat[Ecapa-tdnn (1-to-1 scenario)]{
		\includegraphics[width=0.45\textwidth]{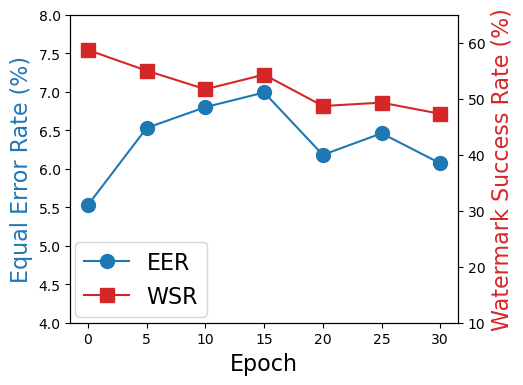}}
   \subfloat[Ecapa-tdnn (1-to-5 scenario)]{
		\includegraphics[width=0.45\textwidth]{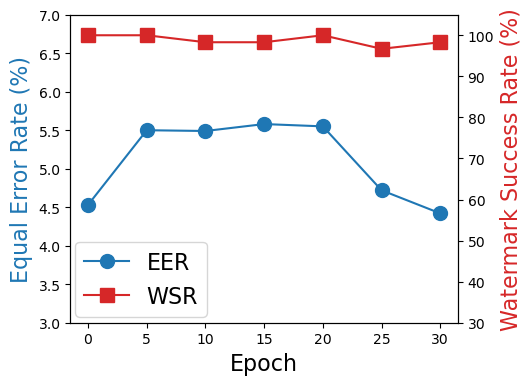}}   \\
   \subfloat[CAM++ (1-to-1 scenario)]{
		\includegraphics[width=0.45\textwidth]{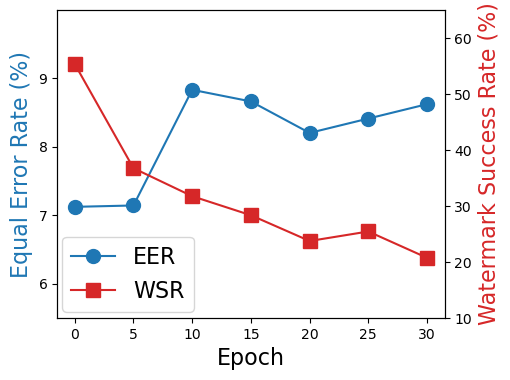}}
   \subfloat[CAM++ (1-to-5 scenario)]{
		\includegraphics[width=0.45\textwidth]{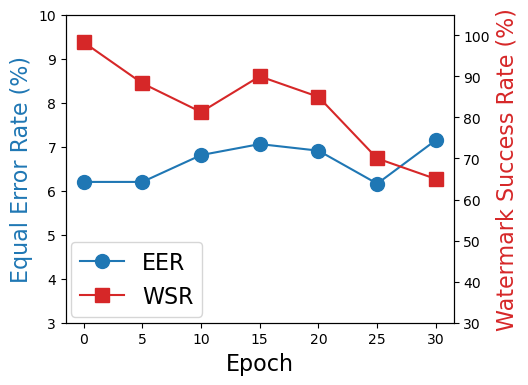}}
        \vspace{-0.5em}
	\caption{Resistance to fine-tuning on the TIMIT dataset.}
     \label{fig:fine-tuning}
    \end{minipage}\hspace{1em}
    \begin{minipage}[t]{0.48\linewidth}
    \centering
     	\subfloat[LSTM (1-to-1 scenario)]{
		\includegraphics[width=0.45\textwidth]{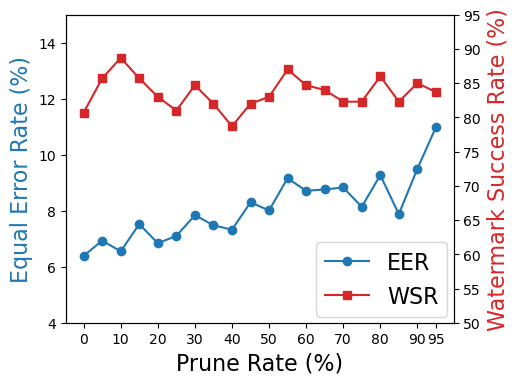}}
	\subfloat[LSTM (1-to-5 scenario)]{
		\includegraphics[width=0.45\textwidth]{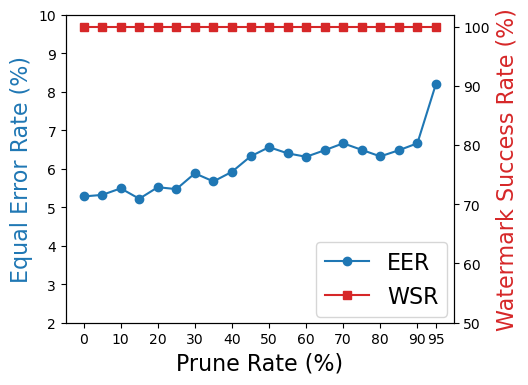}}\\
   \subfloat[Ecapa-tdnn (1-to-1 scenario)]{
		\includegraphics[width=0.45\textwidth]{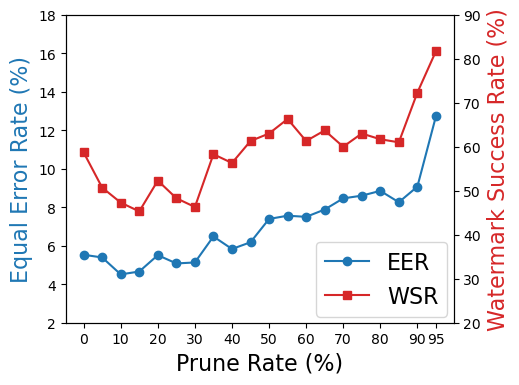}}
    \subfloat[Ecapa-tdnn (1-to-5 scenario)]{
		\includegraphics[width=0.45\textwidth]{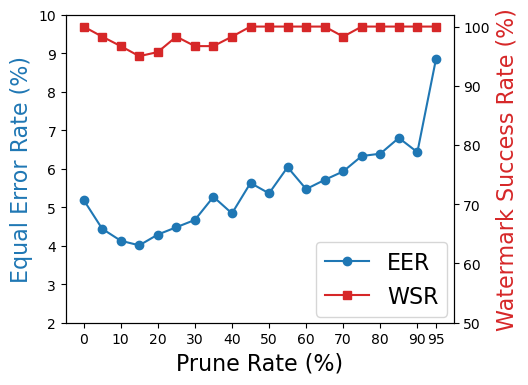}}  \\
    \subfloat[CAM++ (1-to-1 scenario)]{
		\includegraphics[width=0.45\textwidth]{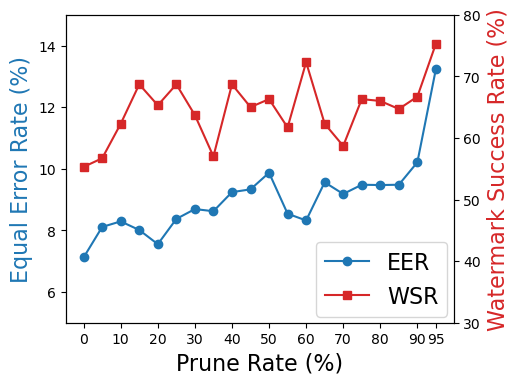}}
    \subfloat[CAM++ (1-to-5 scenario)]{
		\includegraphics[width=0.45\textwidth]{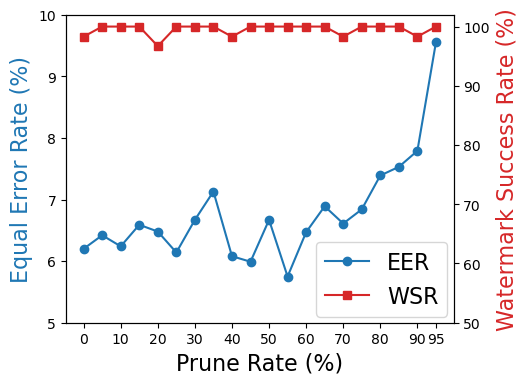}}
        \vspace{-0.5em}
	\caption{Resistance to model pruning on the TIMIT dataset.} 
     \label{fig:pruning}
    \end{minipage}
    \vspace{-1em}
\end{figure*}

\vspace{0.3em}
\noindent \textbf{Resistance to Fine-tuning.} Following the prior work~\cite{liu2017neural}, we adopt 10\% benign samples from the original training set to fine-tune the CBW-watermarked models. As shown in Figure~\ref{fig:fine-tuning}, the WSRs generally decrease as the number of fine-tuning epochs increases, yet our method remains effective to a large extent. Specifically, fine-tuning has relatively limited effects on LSTM and Ecapa-tdnn. For CAM++, its influence is more pronounced, but the WSRs are still high enough for successful ownership verification, especially under the 1-to-5 scenario. These results indicate the resistance of our CBW to the fine-tuning attack.

\vspace{0.3em}
\noindent \textbf{Resistance to Model Pruning.} Following the prior work~\cite{liu2018fine}, we adopt 10\% benign samples from the original training set to prune the feature representations (\ie, embeddings) of the watermarked models, with the pruning rate varying from 0\% to 95\%. As shown in Figure~\ref{fig:pruning}, the EERs increase significantly with the pruning rate, whereas the maximum drop of the WSRs is less than 10\%, \ie, model pruning has a limited impact on the watermark itself. These results suggest that our CBW is resistant to model pruning to a large extent.

\begin{table}[!t]
\caption{Resistance of our CBW to data augmentation on the TIMIT dataset.}
\vspace{-0.5em}
\centering
\scalebox{0.9}{
\begin{tabular}{c|c|c|ccc}
\toprule
Model$\downarrow$        & Scenario$\downarrow$  & Method$\rightarrow$ & Attack & SpecAugment & VD \\ \hline
\multirow{4}{*}{LSTM}        & \multirow{2}{*}{1-to-1} & EER (\%)             & 6.4 & 6.9      & 6.3        \\
                             &                         & WSR (\%)             & 80.7  & 78.7       & 73.7         \\ \cline{2-6} 
                             & \multirow{2}{*}{1-to-5} & EER (\%)             & 5.2 & 5.7      & 5.4         \\
                             &                         & WSR (\%)             & 100.0      & 100.0           & 100.0            \\ \hline
\multirow{4}{*}{Ecapa-tdnn} & \multirow{2}{*}{1-to-1} & EER (\%)             & 5.5 & 5.0      & 5.3        \\
                             &                         & WSR (\%)             & 58.7  & 54.3       & 57.7         \\ \cline{2-6} 
                             & \multirow{2}{*}{1-to-5} & EER (\%)             & 4.5 & 4.1      & 4.4         \\
                             &                         & WSR (\%)             & 100.0      & 93.3       & 100.0             \\ \hline
\multirow{4}{*}{CAM++}       & \multirow{2}{*}{1-to-1} & EER (\%)             & 7.1 & 5.8      & 5.9        \\
                             &                         & WSR (\%)             & 55.3  & 50.3       & 20.6         \\ \cline{2-6} 
                             & \multirow{2}{*}{1-to-5} & EER (\%)             & 6.2  & 4.8       & 4.7        \\
                             &                         & WSR (\%)             & 98.3  & 93.3       & 71.3         \\ \bottomrule
\end{tabular}
}
\label{tab:data augmention}
\vspace{-0.8em}
\end{table}

\vspace{0.3em}
\noindent \textbf{Resistance to Data Augmentation.} Data augmentation is a widely applied technique that generates additional samples for training and might therefore hinder the learning of dataset watermarks. We hereby validate whether it can erase the CBW watermark. Specifically, we exploit the most classical augmentation method for speech, \ie, SpecAugment~\cite{park2019specaugment}, where the time warp parameter, the maximum width of each frequency mask, the maximum width of each time mask, the number of frequency masks, the number of time masks, and the padding value are set to 5, 3, 20, 1, 1, and 0, respectively, restricting the time mask ratio to below 0.2. We also perturb the samples with a small volume disturbance (denoted as `VD') ranging from $-0.5$dB to $0.5$dB for reference. As shown in Table~\ref{tab:data augmention}, both SpecAugment and VD have a limited impact on WSR in most cases and even decrease the EERs in many cases. The most affected case is CAM++ under VD in the 1-to-1 scenario, whose WSR drops to 20.6\%; even so, the corresponding p-value remains far below $\alpha$, because the paired $t$-test draws on the score magnitudes beyond the binary win indicators, retaining power in regimes where the WSR-based certificate in Eq.~(\ref{eq.thm1}) is conservative. These results suggest the resistance of our CBW to classical data augmentation methods.

\begin{table*}[!t]
\begin{minipage}[c]{0.49\textwidth}
\centering
\captionsetup{font=small}
\caption{EER (\%) with different models in 1-to-1 scenario.}
\vspace{-0.5em}
\scalebox{1}{
\begin{tabular}{cc|ccc}
\toprule
                                                   &             & \multicolumn{3}{c}{Target Model} \\ \cline{3-5} 
                                                   &             & LSTM     & Ecapa-tdnn  & CAM++   \\ \hline
\multicolumn{1}{c|}{\multirow{3}{*}{\begin{tabular}[c]{@{}c@{}}Source\\ Model\end{tabular}}} & LSTM        & 6.4   & 6.5      & 8.8  \\
\multicolumn{1}{c|}{}                              & Ecapa-tdnn & 7.2   & 5.5      & 8.3  \\
\multicolumn{1}{c|}{}                              & CAM++       & 6.5   & 6.6      & 7.1  \\ 
\bottomrule
\end{tabular}}
\label{tab:trans-1}    
\end{minipage}\hspace{1em}  
\begin{minipage}[c]{0.49\textwidth}
\captionsetup{font=small}
\caption{WSR (\%) with different models in 1-to-1 scenario.}
\vspace{-0.5em}
\centering
\scalebox{1}{
\begin{tabular}{cc|ccc}
\toprule
\multicolumn{2}{c|}{\multirow{2}{*}{}}                                                                     & \multicolumn{3}{c}{Target Model} \\ \cline{3-5} 
\multicolumn{2}{c|}{}                                                                                      & LSTM    & Ecapa-tdnn   & CAM++   \\ \hline
\multicolumn{1}{c|}{\multirow{3}{*}{\begin{tabular}[c]{@{}c@{}}Source\\ Model\end{tabular}}} & LSTM        & 80.7   & 58.3         & 53.3   \\
\multicolumn{1}{c|}{}                                                                        & Ecapa-tdnn & 73.0    & 58.7         & 58.3   \\
\multicolumn{1}{c|}{}                                                                        & CAM++       & 69.3    & 56.7         & 55.3   \\ 
\bottomrule
\end{tabular}}
\label{tab:trans-2}
\end{minipage}
\end{table*}

\begin{table*}[!t]
\begin{minipage}[c]{0.49\textwidth}
\centering
\captionsetup{font=small}
\caption{EER (\%) with different models in 1-to-5 scenario.}
\vspace{-0.5em}
\scalebox{1}{
\begin{tabular}{cc|ccc}
\toprule
\multicolumn{2}{c|}{\multirow{2}{*}{}}                                                                     & \multicolumn{3}{c}{Target Model} \\ \cline{3-5} 
\multicolumn{2}{c|}{}                                                                                      & LSTM     & Ecapa-tdnn  & CAM++   \\ \hline
\multicolumn{1}{c|}{\multirow{3}{*}{\begin{tabular}[c]{@{}c@{}}Source\\ Model\end{tabular}}} & LSTM        & 5.8   & 5.1       & 6.9  \\
\multicolumn{1}{c|}{}                                                                        & Ecapa-tdnn & 5.8   & 4.5       & 6.3  \\
\multicolumn{1}{c|}{}                                                                        & CAM++      & 5.5   & 5.7       & 6.2   \\ 
\bottomrule
\end{tabular}}
\label{tab:trans-3}
\end{minipage}\hspace{1em}  
\begin{minipage}[c]{0.49\textwidth}
\centering
\captionsetup{font=small}
\caption{WSR (\%) with different models in 1-to-5 scenario.}
\vspace{-0.5em}
\centering
\scalebox{1}{
\begin{tabular}{cc|ccc}
\toprule
\multicolumn{2}{c|}{\multirow{2}{*}{}}                                                                     & \multicolumn{3}{c}{Target Model} \\ \cline{3-5} 
\multicolumn{2}{c|}{}                                                                                      & LSTM    & Ecapa-tdnn   & CAM++   \\ \hline
\multicolumn{1}{c|}{\multirow{3}{*}{\begin{tabular}[c]{@{}c@{}}Source\\ Model\end{tabular}}} & LSTM        & 100.0       & 100.0              & 96.7   \\
\multicolumn{1}{c|}{}                                                                        & Ecapa-tdnn & 100.0        & 100.0              & 96.7   \\
\multicolumn{1}{c|}{}                                                                        & CAM++      & 98.3   & 100.0              & 98.3   \\ 
\bottomrule
\end{tabular}}
\label{tab:trans-4}
\end{minipage}
\vspace{-0.8em}
\end{table*}

\section{Model Transferability of Our CBW}
\label{app:transfer}

As mentioned in Section~\ref{sec:watermarking}, our CBW requires a surrogate feature extractor to obtain the representation of each speaker. Following the classical setting of similar works~\cite{li2022untargeted,guo2023domain,wei2024pointncbw}, we have so far reported the results where the adversary adopts the same model structure as the surrogate in unauthorized training. However, the adversary may adopt a different structure, since the dataset owner usually has no information about the training process in practice. As such, we hereby evaluate the model transferability of our CBW, \ie, whether it remains effective when the surrogate model (the source model in the tables) differs from the target model trained by the adversary.

Specifically, we conduct experiments on the TIMIT dataset, where all settings except the model structures are the same as those in Section~\ref{settings}. As shown in Table~\ref{tab:trans-1}-\ref{tab:trans-4}, our CBW remains effective under all source-target pairs, as anticipated in Section~\ref{sec:watermarking}, although the WSRs fluctuate moderately across architecture pairs. One reading through Proposition~\ref{prop:coverage} is that the anchors are computed from the surrogate's representations, and what the target model learns determines how tightly each trigger lands around its anchor (\ie, the magnitude of $\delta$); the moderate fluctuations are thus consistent with this dependence on the alignment between the two embedding geometries. We also note that in the 1-to-5 scenario the WSRs stay above 96\% for all pairs, consistent with the amplification effect of the enrollment size characterized in Proposition~\ref{prop:coverage}. In other words, our CBW is transferable across different model structures.

\section{Computational Complexity Analysis}
\label{app:complexity}

In this appendix, we analyze the computational complexity of our CBW-based dataset ownership verification.

\vspace{0.3em}
\noindent \textbf{The Complexity of Dataset Watermarking.} Let $n, t, K, d$ denote the number of samples in the training set, the number of iterations for clustering, the number of clusters, and the dimension of the feature representation, respectively. Our CBW first extracts the representations of all samples and then performs clustering, whose complexities are $\mathcal{O}(n)$ and $\mathcal{O}(t \cdot K \cdot n \cdot d)$, respectively. After that, the dataset owner obtains the cluster of each speaker and implants the corresponding trigger patterns, whose complexity is $\mathcal{O}(K)$. As such, the overall computational complexity of dataset watermarking is $\mathcal{O}(n + t \cdot K \cdot n \cdot d + K)$, dominated by the clustering step.

\vspace{0.3em}
\noindent \textbf{The Complexity of Ownership Verification.} In this stage, the dataset owner queries the suspicious model and conducts the hypothesis test. Recall that $K$ is also the number of triggers, so the audit issues $K$ queries per trial and $m \cdot K$ queries in total. Considering the 1-to-$N$ scenario, each trial computes the similarities between the queries and the $N$ enrolled voiceprints and takes the maximum, or counts the acceptances in the decision-only setting, so the computational complexity is $\mathcal{O}(m \cdot K \cdot N)$. This accounting also quantifies the overhead behind the trade-offs on $K$ and $m$ discussed in Appendix~\ref{app:ablation}.

In particular, the dataset owner can further accelerate both stages by processing samples in a batch manner.

\end{document}